\documentclass[aps,prd,unsortedaddress,superscriptaddress,showpacs,nofootinbib,twocolumn,10pt]{revtex4-2} 
\usepackage{natbib}
\usepackage[utf8]{inputenc}
\usepackage{array}
\usepackage{multirow}
\usepackage{graphicx,color}
\usepackage{relsize}
\usepackage{slashed}
\usepackage[normalem]{ulem}
\usepackage{tabu}
\usepackage{rotating}
\usepackage{bigstrut}
\usepackage{makecell}
\usepackage[dvipsnames]{xcolor}
\usepackage[colorlinks=true, linkcolor=blue, citecolor=Green, urlcolor=blue]{hyperref}
\usepackage{amsmath}
\usepackage{amssymb,bm}
\usepackage{amsthm}
\usepackage{mathrsfs}
\usepackage{array}
\usepackage[all]{xy}
\usepackage{euscript}
\usepackage{enumerate}
\usepackage{mathtools}
\usepackage{soul}
\usepackage{orcidlink}

\allowdisplaybreaks

\graphicspath{{Figs/}} 

\newcommand{\itp}{\affiliation{Institute of Theoretical Physics,
Chinese Academy of Sciences, Beijing 100190, China}}
\newcommand{\ucas}{\affiliation{School of Physical Sciences, 
University of Chinese Academy of Sciences, Beijing 100049, 
China}}

\newcommand{\bonn}{\affiliation{Helmholtz Institut f\"{u}r Strahlen- und Kernphysik and Cluster of Excellence ``Color meets Flavor'', Universit\"{a}t Bonn, D-53115 Bonn, Germany}}
\newcommand{\bonnbethe}{\affiliation{Bethe Center for Theoretical Physics, Universit\"{a}t Bonn, D-53115 Bonn, Germany}}

\newcommand{\scnt}{\affiliation{Southern Center for Nuclear-Science Theory (SCNT), Institute of Modern Physics,\\ 
Chinese Academy of Sciences, Huizhou 516000, China}}

\newcommand{\julich}{\affiliation{Institute for Advanced Simulation (IAS-4) and Cluster of Excellence ``Color meets Flavor'', Forschungszentrum J\"ulich, D-52425 J\"ulich, Germany}}

\newcommand{\AIBX}{A_{\rm IB}^X}
\newcommand{\AIB}{A_{\mathrm{IB}}}

\newcommand{\wc}{W_{c1}}

\begin{document}

\title{Amplified Isospin Breaking from Coupled-Channel Dynamics Near Threshold}

\author{Xiang-Kun Dong\orcidlink{0000-0001-6392-7143}}
 \email{xiangkun@hiskp.uni-bonn.de}
\bonn\bonnbethe

\author{Feng-Kun~Guo\orcidlink{0000-0002-2919-2064}}
\email{fkguo@itp.ac.cn}
\itp\ucas\scnt

\author{Christoph Hanhart\orcidlink{0000-0002-3509-2473}}\email{c.hanhart@fz-juelich.de}
\julich

\author{Teng Ji\orcidlink{0000-0003-0366-1042}}
\email{teng@hiskp.uni-bonn.de}
\bonn\bonnbethe

\author{Ulf-G. Mei{\ss}ner\orcidlink{0000-0003-1254-442X}}\email{meissner@hiskp.uni-bonn.de}
\bonn\bonnbethe\julich


\begin{abstract}
We identify the dynamical origin of amplified isospin breaking in near-threshold states. Using a pole-residue-based measure defined directly from the pole couplings, rather than from decay observables that are also affected by different final-state kinematics, we show that threshold proximity of the pole alone is insufficient: large isospin breaking requires a sizable interaction in the companion isospin channel. In particular, the isospin-violating pole-coupling ratio vanishes when the companion-channel interaction is switched off, even at finite threshold splitting. Applied to the $X(3872)$, the observed large isospin breaking points to a nearby partner pole, $W_{c1}$, as predicted in previous studies. As a concrete realization of the previously discussed leading/shadow-pole interchange, we find that the $W_{c1}$ pole, which predominantly affects
the line shapes near the charged threshold in the
physical case, does not evolve into the pure $I=1$
eigenstate in the isospin limit; instead, that eigenstate
is continuously connected to a more distant shadow pole.
\end{abstract}

\maketitle
\section{Introduction}

Near-threshold coupled-channel dynamics has become a central theme in modern hadron spectroscopy, since nearby thresholds can distort line shapes, shift pole positions and play a crucial role in the experimental visibility of states~\cite{Guo:2017jvc,Guo:2014iya,Dong:2020hxe,Baru:2021ldu,Zhang:2024qkg,Sone:2024nfj}. The $X(3872)$, a typical example in this class, was first discovered by Belle and later confirmed by CDF and D\O~\cite{D0:2004zmu,Belle:2003nnu,CDF:2003cab}. It is located within sub-MeV proximity to the $D^0\bar D^{*0}$ threshold~\cite{ParticleDataGroup:2026aaa} and couples strongly to this channel~\cite{Belle:2008fma,Li:2019kpj,Braaten:2019ags,BESIII:2020nbj,BESIII:2023hml}, thereby making universal low-energy dynamics unavoidable~\cite{Braaten:2003he,Braaten:2004rn}.
Such properties imply a substantial molecular component, although they do not completely exclude short-distance contributions. Consequently, the $X(3872)$ continues to play a central role in the molecular-versus-compact debate, including hybrid scenarios in which hadronic and short-distance components coexist~\cite{Tornqvist:2004qy,Maiani:2004vq,Swanson:2003tb}. Recent reviews of exotic hadrons with explicit discussions of the $X(3872)$ can be found in Refs.~\cite{Lebed:2016hpi,Hosaka:2016pey,Esposito:2016noz,Olsen:2017bmm,Guo:2017jvc,Ali:2017jda,Guo:2019twa,Brambilla:2019esw,Dong:2021juy,Chen:2022asf,Dai:2026fkg}.

The most striking feature of the $X(3872)$ is the coexistence of a pole located extremely close to threshold~\cite{LHCb:2020xds,BESIII:2023hml,Ji:2025hjw} with pronounced isospin breaking. The neutral and charged $D\bar D^*$ thresholds differ by $\Delta \simeq 8.23~\mathrm{MeV}$~\cite{ParticleDataGroup:2026aaa}, which is small on hadronic scales but large compared to the sub-MeV binding energy $E_b$ typically inferred for the $X(3872)$, with the pole determined to be $\left(-160^{+57}_{-74}-125^{+23}_{-38}\,i\right)$~keV relative to the $D^0\bar D^{*0}$ threshold in Ref.~\cite{Ji:2025hjw}. Experimentally, no charged partner has been established~\cite{BaBar:2004cah,Belle:2011vlx}, favoring an isoscalar assignment. At the same time, sizable isospin violation is observed in its hidden-charm decay pattern, as reflected in the $J/\psi\pi\pi$ and $\chi_{cJ}\pi^0$ modes~\cite{CDF:2005cfq,LHCb:2022jez,BESIII:2019esk,Belle-II:2026xdx}. More quantitatively, after accounting for the $\rho$--$\omega$ mixing and the very different 
$2\pi$ (mediated by the 
$\rho$ with a width of 150~MeV) and $3\pi$ (mediated by the $\omega$ with a width of 8~MeV) phase spaces, our dispersive analyses in Refs.~\cite{Ji:2025hjw, Dias:2024zfh} yield
\begin{align}
 R_X\equiv\left|\frac{g_{XJ/\psi\rho}}{g_{XJ/\psi\omega}}\right|\simeq0.26\pm0.02
 \label{eq:RXexp}
\end{align}
This result is consistent with previous determinations~\cite{Suzuki:2005ha,Braaten:2005ai,Hanhart:2011tn,LHCb:2022jez,Wang:2022vjm} and is substantially larger than the natural percent-level short-distance isospin breaking expected from the light-quark mass difference and electromagnetic effects. Here and in the following, we refer to isospin breaking as sizable when $R_X$, or equivalently the $|\AIB|$ defined below, reaches the $\mathcal O(0.1)$ level or above, i.e., when it is parametrically larger than the natural percent-level short-distance isospin breaking. In the power counting of our effective field theory, such short-distance isospin-breaking operators are therefore treated as subleading, while the leading enhancement is generated by the low-energy $D\bar D^*$ dynamics. Earlier analyses have focused on how the threshold splitting of $D\bar D^*$ can strongly amplify isospin violation in near-threshold systems~\cite{Hanhart:2007yq,Gamermann:2009fv,Gamermann:2009uq,Hidalgo-Duque:2012rqv,Li:2012cs,Takeuchi:2014rsa,Wu:2021udi,Meng:2021kmi}. What has received less attention is how, within this power counting, the resulting isospin breaking at the $X(3872)$ pole depends on the interaction in the companion $I=1$ channel. 

In this work, we adopt 
a minimal two-channel framework to demonstrate that for a near-threshold molecular state, a large threshold splitting alone does not necessarily imply large isospin breaking and that distinguishing these two effects is a central issue. More precisely, using the pole-coupling ratio employed in
Refs.~\cite{Meng:2021kmi,Zhang:2024fxy} that quantifies the isovector admixture of an isoscalar-dominated pole, we show that threshold splitting provides only the seed of isospin breaking, while large effects require sizable interactions in an isospin sector different from the main one. This mechanism can be straightforwardly generalized to other isospin sectors. 
Applied to the $X(3872)$, the observed large isospin breaking then points to the presence of a nearby $I=1$ partner pole $\wc$, as predicted in our previous studies~\cite{Zhang:2024fxy,Ji:2025hjw}. The present analysis provides an explicit dynamical connection
between the enhanced isospin breaking and the interaction
responsible for the companion pole.

\section{Minimal Coupled-Channel Framework}\label{sec:Frame}
We employ a minimal two-channel framework 
that allows us to investigate isospin breaking at the level of
the pole couplings, independently of channel-dependent kinematic effects that enter specific decay observables. 
Although the mechanism discussed here is very general, for illustration in what follows we focus on the 
example of the $X(3872)$.
Its partial widths into $J/\psi\rho$ and $J/\psi\omega$ are not determined by the corresponding couplings alone, but involve integrals over the vector-meson line shapes,
\begin{equation}
 \Gamma_{X\to J/\psi \mathcal V}
\propto
|g_{XJ/\psi \mathcal V}|^2
\int ds\,\rho_{\mathcal V}(s)\,\Phi_{J/\psi \mathcal V}(s)
\end{equation}
for ${\mathcal V}=\rho,\omega$, where $\rho_{\mathcal V}(s)$ denotes the spectral function of the unstable vector meson and $\Phi_{J/\psi \mathcal V}(s)$ the corresponding phase-space factor. By working directly with pole couplings, our analysis removes these channel-specific factors. We emphasize that $\AIB$ is the ratio of the isovector to isoscalar pole couplings. The isospin decomposition of the full wave function additionally involves the neutral- and charged-channel propagators, whose different threshold dependence is not captured by this coupling ratio alone.

The essential ingredients are therefore the two nearby $D\bar D^*$ thresholds and the interactions in the isospin channels. Additional inelastic channels determine observable widths and decay distributions and are indispensable for a quantitative description of specific final states, but they do not change the basic mechanism by which the charged--neutral threshold splitting modifies the isospin composition of the near-threshold pole. The present setup thus isolates this two-threshold pole dynamics from the additional ingredients required to describe particular decay observables.

The basis in which we formulate the equations
is $(|n\rangle, |c\rangle)^T$, where
$|n\rangle=|D^0\bar D^{*0}\rangle$ and $|c\rangle=|D^+D^{*-}\rangle$ denote the neutral and charged channels, respectively\footnote{Charge conjugation is imposed implicitly for fixed $C$-parity and is irrelevant for the present discussion.}. The corresponding thresholds are $\Sigma_n$ and $\Sigma_c$ and their splitting $\Delta\equiv\Sigma_c-\Sigma_n$. The interaction potential can be written as~\cite{Hidalgo-Duque:2012rqv}
\begin{align}
V=
\frac{1}{2}\begin{pmatrix}
C_0+C_1 & C_0-C_1\\
C_0-C_1 & C_0+C_1
\end{pmatrix},\label{eq:V2}
\end{align}
where the leading contact interactions, denoted by $C_0$ and $C_1$, correspond to the isoscalar and isovector interaction strengths, respectively.
Since we are focusing on the near-threshold region, momentum-dependent terms are omitted. Direct isospin breaking in the short-range interaction would introduce corrections, making the diagonal terms unequal in Eq.~\eqref{eq:V2}. 
Such corrections arise from the up–down quark mass difference and electromagnetic effects, which scale as 
\begin{align}
 \mathcal{O}(\delta_I)\sim \mathcal{O}\left(\frac{m_d-m_u}{\Lambda_{\rm QCD}}\right)\sim \mathcal{O}(\alpha_{\rm QED})\sim 10^{-2},
 \label{eq:deltaI}
\end{align}
and are therefore subleading and neglected in this study.

The propagator is given by
\begin{align}
 G(E)=\mathrm{diag}(G_n(E),G_c(E)),
\end{align}
where
\begin{equation}
G_{n,c}(E)=R(\Lambda)-i\frac{\mu_{n,c}}{2\pi}k_{n,c}
\label{eq:G}
\end{equation}
in the nonrelativistic approximation, where $\mu_{n,c}$ is the reduced mass in the corresponding channel.
The momenta $k_{n,c}$ are analytically continued to the complex plane, which defines two Riemann sheets (RSs) for each channel: RS$_+$ for ${\rm Im}(k)> 0$ and RS$_-$ for ${\rm Im}(k)<0$. For the coupled two-channel system, the sheet is denoted by RS$_{ab}$, where $a,b\in\{+,-\}$ label the neutral and charged sheets, respectively. The channel-independent subtraction $R(\Lambda)$ is a leading-order prescription for the analytic short-distance part of the loop functions. For unequal channel masses, a general subtraction contains a common isospin-conserving part, absorbed into $C_{0}$ and $C_{1}$, and a channel-dependent part that induces off-diagonal mixing in the isospin basis. This part formally calls for an isospin-breaking counterterm.
However, 
 these isospin violating parts are not enhanced and accordingly expected to be of natural order, $\delta_I$ provided
 in Eq.~\eqref{eq:deltaI}, and are therefore subleading.
  In contrast, the nonanalytic dependence on the split thresholds is retained exactly in the low-energy amplitude.

The scattering amplitude is written as
\begin{align}
T(E)=\bigl[V^{-1}-G(E)\bigr]^{-1}.
\end{align}
The cutoff dependence of the propagators can be absorbed into the contact interactions via
\begin{align}
C_I^{\rm R}\equiv \left(C_I^{-1}-R(\Lambda)\right)^{-1},\qquad I=0,1,
\end{align}
where $C_I^{\rm R}$ denotes the renormalized interaction strength in the isospin-$I$ channel.
They are equal to $(2\pi/\mu)a_0^{(I)}$ with $\mu$ the reduced mass and $a_0^{(I)}$ the scattering lengths in the isospin limit.
Rotating to the isospin basis, the potential becomes diagonal while all mixing arises from the loop sector. The inverse amplitude reads
\begin{align}
T_I^{-1}(E)=
\begin{pmatrix}
\left(C_0^{\rm R}\right)^{-1}-\bar G & -\delta G\\
-\delta G & {\left(C_1^{\rm R}\right)}^{-1}-\bar G
\end{pmatrix},
\end{align}
with
\begin{equation}
\begin{aligned}
\bar G&=-\frac{i}{2\pi}\frac{\mu_n k_n+\mu_c k_c}{2},\\
\delta G&=-\frac{i}{2\pi}\frac{\mu_n k_n-\mu_c k_c}{2}~.
\end{aligned}
\end{equation}
The off-diagonal terms in the inverse $T$-matrix, $\delta G$, proportional to the difference between the phase-space factors of the two channels, show a very specific energy dependence as discussed in Ref.~\cite{Achasov:1979xc} for the example of $a_0(980)$-$f_0(980)$ mixing. The predictions for the emerging peculiar line shapes for $J/\psi\to \phi \pi^0\eta$ from the more refined study in Ref.~\cite{Hanhart:2007bd} were confirmed experimentally by BESIII~\cite{BESIII:2010dhc}.

Near a pole of the scattering amplitude, one has the Laurent expansion,
\begin{align}
T_{ij}(E)\to\frac{g_i g_j}{E-E_{\rm pole}}+\text{regular~terms},\label{eq:residual}
\end{align}
which defines the channel couplings $g_n,g_c$. In our minimal two-channel model, the isoscalar and isovector sectors are formally symmetric. We therefore focus, without loss of generality, on the case of a near-threshold pole generated by the isoscalar interaction, namely, one that evolves into a pure isoscalar pole as the isovector interaction is gradually switched off. The corresponding isovector case follows by interchanging the relevant isospin labels and yields symmetric conclusions. The isospin-breaking measure of this pole is then defined by the ratio of the isovector to the isoscalar coupling,
\begin{align}
\AIB\equiv\frac{g_n-g_c}{g_n+g_c}
&=\frac{\delta G(E_{\rm pole})}{(C_1^{\rm R})^{-1}-\bar G(E_{\rm pole})}.
\label{eq:AIB}
\end{align}
The first equality is the isospin-breaking measure used in Refs.~\cite{Meng:2021kmi,Zhang:2024fxy}, whereas the second equality is derived here. It captures the point of the mechanism: $\delta G$ provides the seed of isospin breaking, while the denominator encodes its amplification through the $I=1$ interaction. Thus, at a fixed pole position and fixed thresholds, the magnitude
of the coupling ratio is controlled by the companion-channel
interaction. A small binding energy alone therefore does not
fix the isospin-violating pole-coupling ratio.

Being defined as the ratio of the pole couplings in the isovector and isoscalar channels, $\AIB$ is a quantity determined solely by the pole properties. Its relation to decay observables depends, in general, on the short-distance transition operators and on the kinematic factors associated with the final states. 
We denote the short-distance transitions $D\bar D^*\to J/\psi\mathcal V$ by $u_{a\mathcal V}$, with $a=n,c$ and $\mathcal V=\rho,\omega$. They can be related to each other by treating them as a quartet under light-flavor U(2), which can be further decomposed into a triplet ($\rho$) and a singlet ($\omega$).
There are two distinct corrections to the relations among the members of the U(2) quartet:
first, genuine short-distance isospin breaking is expected to be of its natural percent-level size (see discussion around Eq.~\eqref{eq:deltaI}),
\begin{align}
u_{n\rho}
&=
-u_{c\rho}\left(1+\epsilon_\rho\right),
&
u_{n\omega}
&=
u_{c\omega}\left(1+\epsilon_\omega\right)\label{eq:u-deltaI}
\end{align}
 with $\epsilon_\rho,\epsilon_\omega\sim\mathcal O(\delta_I)$, which are therefore subleading in the power counting adopted here;
second, corrections due to the violation of Okubo-Zweig-Iizuka (OZI) rules~\cite{Okubo:1963fa,Zweig:1964jf,Iizuka:1966fk} (for more details, see the discussion in Appendix~\ref{app:short-distance-flavor} and Ref.~\cite{Meissner:2000bc}) lead to
\begin{equation}
\begin{aligned}
u_{n\rho}=
u_{n\omega}
\left(
1+\eta_n
\right),\quad
u_{c\rho}=-
u_{c\omega}
\left(
1+\eta_c
\right),
\label{eq:OZI-relation}
\end{aligned}
\end{equation}
with $\eta_n,\eta_c\sim\mathcal O(\delta_{\rm OZI})$ parameterizing the relative size of the OZI-suppressed contribution. A naive estimate of its size is provided by the 
small deviation of the phenomenological $\omega$--$\phi$ mixing angle from ideal mixing, $\theta_V\simeq36.5^\circ$ compared with $\theta_V^{\rm ideal}\simeq35.3^\circ$~\cite{ParticleDataGroup:2026aaa}, corresponding to a relative deviation of only a few percent. This is also consistent with phenomenological analyses in which the $\omega$ coupling was either fitted independently or fixed by the corresponding flavor relation, with mutually compatible results~\cite{Daub:2015xja,Heuser:2025mnk,Heuser:2026glv}. However, the size of OZI violation is known to be strongly process dependent and can be substantially larger in other hadronic reactions~\cite{Lipkin:1996ny,Meissner:2000bc,Sibirtsev:2005nq,Nomokonov:2002jb}. We therefore adopt
$
\delta_{\rm OZI}\sim0.1$ 
as a conservative representative amplitude-level benchmark for the present analysis, rather than as a universal estimate or strict upper bound.

The coupling ratio extracted after removing the final-state spectral and phase-space effects is then 
\begin{align}
R_X
&=
\left|
\frac{
g_n u_{n\rho}+g_c u_{c\rho}
}{
g_n u_{n\omega}+g_c u_{c\omega}
}
\right|
\nonumber\\
&=\left|
\frac{
g_n (1+\epsilon_I)-g_c
}{
g_n(1+\epsilon_I)+g_c
}(1+\eta)
\right|\notag\\
&=
\left|\AIB (1+\eta)
+\frac{1-\AIB^2}{2}\epsilon_I\right|+\mathcal O(\epsilon_I^2,\eta^2,\epsilon_I\eta),
\label{eq:RX-AIB}
\end{align}
where we have assumed $\epsilon_\rho=\epsilon_\omega=\epsilon_I$ and $\eta_n=\eta_c=\eta$.
Thus, within the present power counting, the experimentally extracted coupling ratio $R_X$ directly probes the isospin-breaking measure $|\AIB|$ determined from pole parameters, up to OZI-suppressed corrections of relative size $\mathcal O(\delta_{\rm OZI})$ and ordinary short-distance isospin-breaking effects of order $\delta_I$. 
Specifically, under this common-correction assumption, corrections from the direct short-distance isospin breaking vanish as $\AIB\to\pm1$, leaving the OZI correction dominant at this order. On the other hand, if $\AIB$ would vanish, direct short-distance isospin breaking
would provide the leading contribution to the decay-coupling ratio, which, however,
would be at odds with
Eq.~\eqref{eq:deltaI} given the smallness of $\delta_I$. 

\section{Amplification Mechanism}\label{sec:amplification}

\begin{figure}[t]
 \centering
 \includegraphics[width=\linewidth]{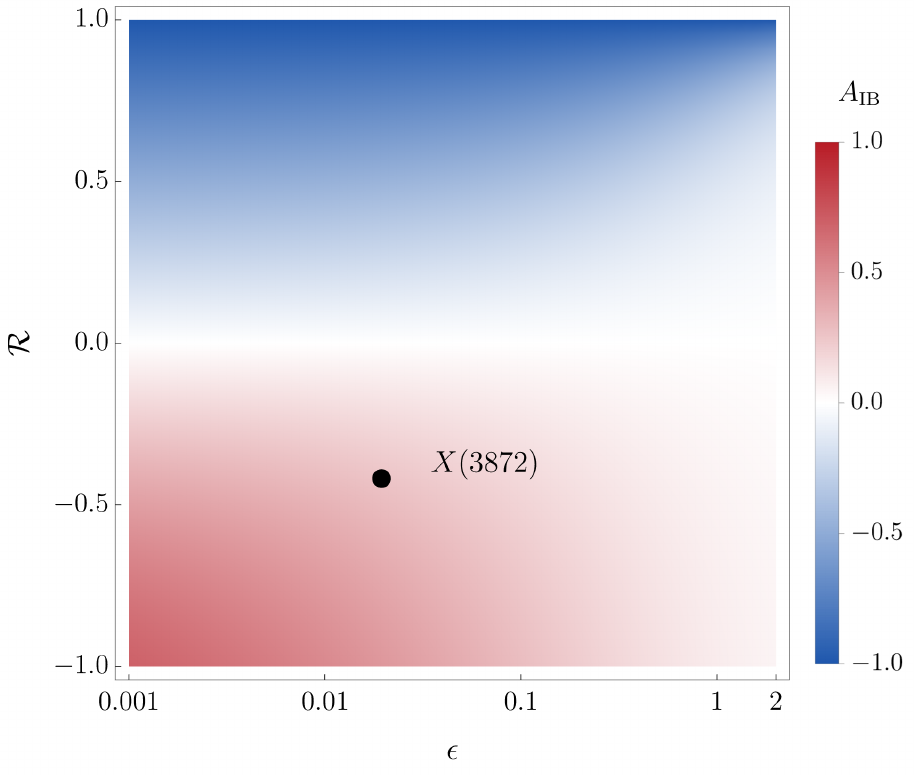}
 \caption{Phase map of $\AIB$ for the bound-state scenario, with $\epsilon$ and $\mathcal R$ defined in Eq.~\eqref{eq:epsilon}. The black dot indicates the $X(3872)$ case, with its parameters taken from Ref.~\cite{Ji:2025hjw}.}
 \label{fig:phase}
\end{figure}

It is evident from Eq.~\eqref{eq:AIB} that threshold splitting alone is insufficient to generate significant isospin breaking, since $\AIB\to 0$ when the renormalized $I=1$ interaction vanishes, $C_1^{\rm R}\to 0$, even for $\Sigma_n\neq\Sigma_c$. For illustration, we consider a pole on the real axis of RS$_{++}$ below $\Sigma_n$, with a binding energy
\begin{align}
E_b\equiv\Sigma_n-E_{\rm pole},
\end{align}
which is the scenario for the $X(3872)$ extracted from data in Ref.~\cite{Ji:2025hjw}.
Here $E_\text{pole}$ is the real part of the pole position in the complex energy plane.
To make the parametric dependence explicit, we introduce two dimensionless variables
\begin{align}
\epsilon\equiv E_b/\Delta,\qquad
\mathcal R \equiv C_1^{\rm R}/C_0^{\rm R},\label{eq:epsilon}
\end{align}
where $\epsilon$ measures the proximity of the real part of the pole to the lower threshold and $\mathcal R$ encodes the relative strength of the renormalized interactions. The isospin-breaking ratio then becomes
\begin{align}
\AIB 
=
\frac{-2 \mathcal R}{K^{2}(1- \mathcal R)+\sqrt{K^{4}(1-{\mathcal R})^{2}+4\mathcal R}},
\end{align}
with $K\equiv\sqrt{1+\epsilon}+\sqrt \epsilon$. The dependence of $\AIB$ on $\epsilon$ and $\mathcal R$,
shown in Fig.~\ref{fig:phase}, illustrates the interplay between
the threshold splitting and the companion-channel interaction.
In general, $|\AIB|$ tends to increase as $\epsilon$ decreases
and $|\mathcal R|$ increases, whereas $\AIB$ vanishes as $\mathcal R\to0$,
even at finite threshold splitting.
For the $X(3872)$, the values $E_b\simeq160~\mathrm{keV}$ and
$R_X=0.26$ from Ref.~\cite{Ji:2025hjw} correspond to
$(\epsilon,\mathcal R)=(0.02,-0.41)$, illustrating this
shallow-binding regime.

\section{From Amplification to Partner Poles}\label{sec:IB-pole}

We have shown that nonzero isospin breaking, $\AIB\neq 0$, requires a nonvanishing isovector interaction. More importantly, a large $\AIB$ can only be obtained when
\begin{equation}
\left|
1/C_1^{\rm R}-\bar G(E_X)
\right| \notag
\end{equation}
is small at the $X(3872)$ pole position, i.e., not much larger than $\delta G(E_X)\simeq\mu_c^{3/2}\sqrt{2\Delta}/(4\pi)$. 
 Thus, the amplification of isospin breaking is not merely a kinematic consequence of the charged--neutral threshold splitting. It also requires a sufficiently strong isovector interaction. 
Once this condition is satisfied, the same interaction that amplifies the isospin-breaking component of the $X(3872)$ produces an additional nearby pole in the isovector channel: 
in the limit $E_b\ll\Delta$, the loop-function difference is governed by the momentum scale $k_\Delta=\sqrt{2\mu\Delta}$.
The characteristic momentum scale associated with the isovector
interaction is set by its inverse scattering length,
\begin{align}
|\gamma_1|
\equiv \left|\frac{1}{a_0^{(1)}}\right|
&= \frac{k_\Delta}{2}
\left|K-\frac{1}{A_{\rm IB}K}\right|
\nonumber\\
&\simeq \frac{k_\Delta}{2}
\left|1-\frac{1}{A_{\rm IB}}\right|,
\qquad \epsilon\ll1,
\end{align}
where the second line gives the leading shallow-binding estimate. Near $A_{\rm IB}=1$, the leading terms cancel and the
finite-binding terms in the first line must be retained.
This scale characterizes the companion-channel interaction and is not a determination of the charged-channel momentum of
the physical $W_{c1}$ pole. In the decoupling limit $A_{\rm IB}=1$, the charged channel
supports a pole with a similar binding energy relative
to its own threshold.

A direct calculation for the $X(3872)$ system,
using the pertinent parameters extracted
from data in Ref.~\cite{Ji:2025hjw}, namely
$E_b=160~\mathrm{keV}$ below the neutral threshold on RS$_{++}$ and $\AIB=0.26$, leads to a nearby $I=1$ pole on RS$_{+-}$ close to the charged threshold at
\begin{equation}
E_{\wc}-\Sigma_c=(1.8+1.2\,i)~\mathrm{MeV},
\end{equation}
consistent with Refs.~\cite{Zhang:2024fxy,Ji:2025hjw}; see also Ref.~\cite{Sadl:2024dbd} for support from lattice calculations. Although the precise pole position depends on the details of the interaction, the qualitative conclusion is robust: once a near-threshold bound state exhibits a large isospin-breaking component, a nearby pole associated with the companion isospin channel and located close to the corresponding threshold is generally implied.

Isovector partners are also expected in compact-tetraquark scenarios, where they arise as members of the same flavor multiplet as the $X(3872)$ and are therefore typically predicted to be nearly degenerate with it (mass splittings are driven by the quark mass differences directly)~\cite{Maiani:2004vq,Maiani:2017kyi,Maiani:2020zhr}. The underlying mechanism is, however, qualitatively different from that discussed here. In a compact-state picture with perturbative coupling to the open-charm continuum, the isoscalar and isovector partners share essentially the same intrinsic resonance dynamics, up to isospin-dependent masses, widths, and decay couplings, and are therefore expected to exhibit broadly similar resonance profiles. In the present scenario, by contrast, the companion poles are generated by the $D\bar D^*$ interaction itself and are tied to the nearby neutral and charged thresholds. Their line shapes therefore depend strongly on the pole positions relative to the corresponding threshold as well as on the associated Riemann sheet.\footnote{As discussed in Refs.~\cite{Zhang:2024fxy,Ji:2022blw}, the location of the $W_{c1}$ pole on a remote Riemann sheet naturally explains why its signal remains indiscernible so far.} Thus, both the location of the partner state and its threshold line shape provide discriminating information on the underlying mechanism.

\begin{figure*}
\centering
\includegraphics[width=\linewidth]{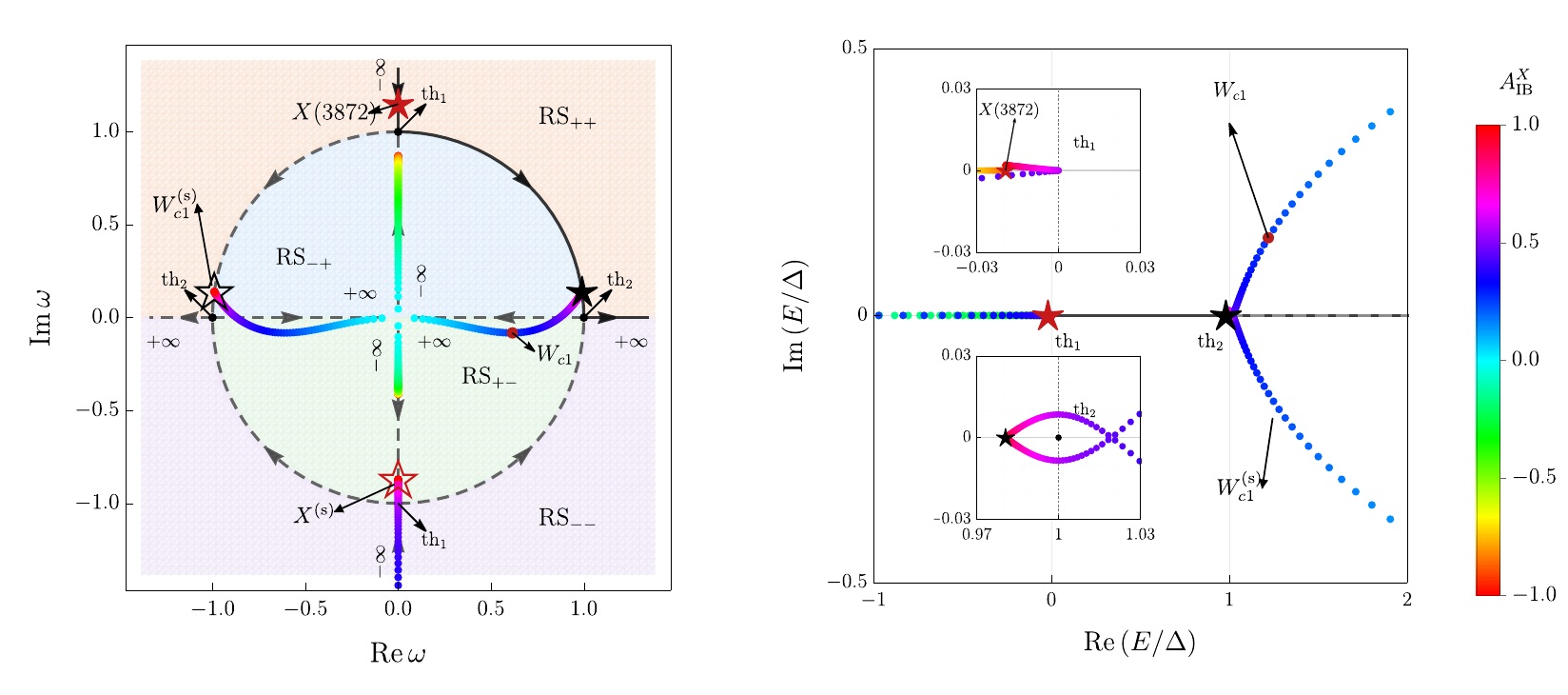}
\caption{{\bf Left}: Pole trajectories in the $\omega$ plane at the fixed $X(3872)$ position, with $E_b=160$~keV, as the isospin-breaking measure $\AIBX$ of the $X(3872)$ varies between $-1$ and $+1$. 
At $\AIBX=1$, the $X$ pole and its shadow are denoted by red filled and open stars, respectively, while the $W_{c1}$ and its shadow are denoted by black filled and open stars, respectively.
The horizontal and vertical axes, together with the unit circle in the $\omega$ plane, correspond to the real energy axes on different Riemann sheets. The solid line segments denote the physical axis. The arrows indicate the positive directions of $E$. {\bf Right}: The same pole trajectories shown in the energy plane. The two inset panels zoom in on the regions around the neutral and charged thresholds. In the zoom-in plot near the $D^0\bar D^{*0}$ threshold, the shadow-pole trajectory of the $X(3872)$ on RS$_{+-}$ and RS$_{--}$ lies on the real axis; it is slightly displaced into the complex plane for visual clarity.} 
\label{fig:x3872_omega_E_plane}
\end{figure*}

\subsection{Pole trajectories as the interaction varies}
To make explicit the connection between large isospin breaking and nearby companion poles,
we vary the isospin-violating parameter
to move
along a curve in the $(C_0^{\rm R},C_1^{\rm R})$ plane, defined by 
the pole location of
the $X(3872)$ being fixed
at its physical value.
In the following, the isospin-breaking measure with respect to the pole of the $X(3872)$ is denoted by $\AIBX$. 

It is useful to describe the trajectories in terms of the uniformizing variable $\omega$~\cite{Kato:1965iee},
\begin{equation}
\begin{aligned}
k_{n,c}&=\sqrt{\frac{\mu \Delta}{2}}\left(\omega\pm\frac{1}{\omega}\right),
\end{aligned}
\end{equation}
which maps the four Riemann sheets of the two-channel problem onto a single complex $\omega$ plane. Throughout this discussion, the sheet labels RS$_{\pm\pm}$ are defined in the particle basis as in Sec.~\ref{sec:Frame}: the first sign specifies the analytic branch of the momentum $k_n$ in the neutral-meson-pair channel, and the second sign specifies that of the momentum $k_c$ in the charged-meson-pair channel. This convention is important. In particular, a mixed sheet such as RS$_{+-}$ or RS$_{-+}$ means that the neutral and charged particle propagators are continued to different branches. 

\begin{figure}[t]
\centering
\includegraphics[width=\linewidth]{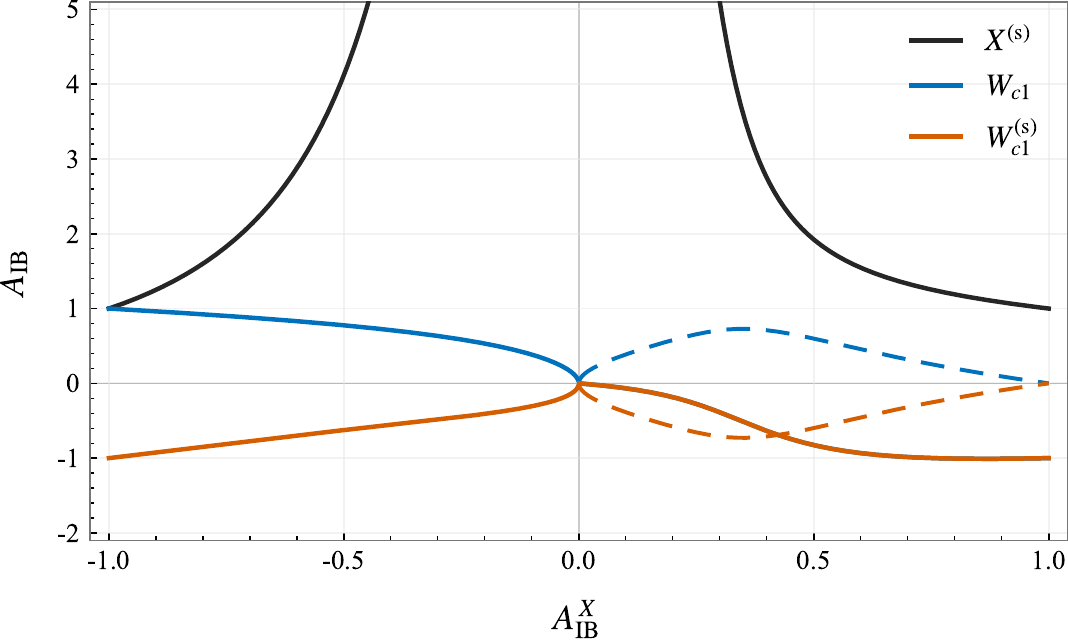}
\caption{Isospin-breaking measures of the other three poles at the fixed $X(3872)$ position, with $E_b=160$ keV, as the isospin-breaking measure of the $X(3872)$, $\AIBX$, is varied. Notice that, in the region with $\AIBX>0$, the real part of $A_{\rm IB}^{W_{c1}}$ (blue solid curve) is hidden beneath that of $A_{\rm IB}^{W_{c1}^{\rm (s)}}$ (orange solid curve); their imaginary parts are shown as dashed curves. For $\AIBX<0$, all poles lie on the real axis, and the corresponding $A_{\rm IB}$ values are purely real.}
\label{fig:AIB-others}
\end{figure}

For the contact interaction considered here, the scattering amplitudes of the two coupled channels give four poles in total~\cite{Zhang:2024qkg}. One of them is the fixed $X(3872)$ pole. 
In Fig.~\ref{fig:x3872_omega_E_plane}, the pole position of the $X(3872)$ that is held fixed is marked by the
filled red star. The remaining three poles are the shadow pole of the $X$, denoted by $X^{\rm(s)}$ and marked by an open red star, and the $W_{c1}$ pole pair. The latter consists of a leading pole, $W_{c1}$ (filled black star), close to the physical region and a shadow pole, $W_{c1}^{\rm(s)}$ (open black star). These two poles form a complex-conjugate pair in the energy plane. All three poles move as the interaction parameters are varied to produce the specified $\AIBX$ while keeping the $X(3872)$ pole fixed.
Figure~\ref{fig:x3872_omega_E_plane} shows the trajectories of the three poles in both the $\omega$ plane and the energy plane as $\AIBX$ is varied. Their isospin structures in terms of $A_{\rm IB}$ along the pole trajectories are displayed in Fig.~\ref{fig:AIB-others}.

The endpoint $\AIBX=1$ provides a starting point for viewing the trajectories. In this limit, $C_0=C_1$ in Eq.~\eqref{eq:V2} and the $X(3872)$ couples only to the neutral channel $D^0\bar D^{*0}$. Correspondingly, the potential matrix becomes proportional to the identity matrix in the particle basis, the off-diagonal transition between the neutral and charged channels vanishes, and the two channels therefore decouple. Since the charged channel feels the same diagonal interaction, it also supports a bound-state pole on RS$_{-+}$, denoted by $W_{c1}$, with almost the same binding energy as that of $X(3872)$, measured from the charged threshold. 
Because the $W_{c1}$ couples only to $D^+D^{*-}$ at this endpoint, its isospin-breaking measure is
$A_{\rm IB}^{W_{c1}}=-1$,
as shown by the blue solid curve in Fig.~\ref{fig:AIB-others}, the real part of which is covered by the orange solid curve corresponding to its shadow in the region of $\AIBX>0$. In the same decoupling limit, each bound-state pole is accompanied by a shadow pole $X^{(s)}$ on a different sheet, which is a generic feature of coupled-channel scattering~\cite{Ross:1963lhp,Dalitz:1963ek,Eden:1963zz,Eden:1964zz,Zhang:2024qkg}. Thus the neutral-channel bound state $X(3872)$ is accompanied by a shadow pole on RS$_{+-}$, while the charged-channel bound state
$W_{c1}$ has its shadow partner $W_{c1}^{\rm (s)}$ at the conjugate position on RS$_{-+}$ as mentioned above. 
At $\AIBX=1$, these shadow poles carry the same channel contents as their corresponding bound-state poles and they differ only by the Riemann sheets on which they reside.

As $\AIBX$ is reduced from unity, the off-diagonal interaction in Eq.~\eqref{eq:V2} is switched on. The neutral and charged channels begin to communicate, while the physical $X(3872)$ pole is kept fixed by construction in this demonstration. The three accompanying poles then reorganize themselves in a highly correlated way.\footnote{The pole trajectories here are obtained by varying $\AIBX$ while the $X(3872)$ pole is kept fixed to its physical value, which is achieved by a correlation between the diagonal and off-diagonal interaction strengths. The picture complements the trajectories shown in Ref.~\cite{Zhang:2024qkg}, where pole trajectories are obtained by varying either the diagonal or the off-diagonal interaction strength at a fixed threshold separation.} The shadow pole $X^{\rm (s)}$ first moves toward the neutral threshold. It then passes through the branch-point structure and migrates from RS$_{+-}$ onto RS$_{--}$. Along this path it remains far from the physical axis, and therefore its direct impact on the observable line shape is limited. Its isospin structure, however, changes dramatically. As shown by the black curve in Fig.~\ref{fig:AIB-others}, the corresponding isospin-breaking measure grows beyond unity and eventually diverges as $\AIBX\to0$. In the normalization used here, this divergence signals the disappearance of the isoscalar projection: the pole becomes purely isovector in this limit. At the same time, its pole position runs to negative infinity along the real axis on RS$_{--}$.

The $W_{c1}$ pole follows a different trajectory. Once the coupling to the lower neutral channel is turned on, the $W_{c1}$ is no longer a stable bound-state pole on the real axis. It acquires an imaginary part and moves into the complex
energy plane on RS$_{-+}$. As $\AIBX$ is lowered, $W_{c1}$ moves through the complex plane and, for $\AIBX\lesssim0.5$, passes through the real axis above the charged threshold and thus moves onto RS$_{+-}$, as shown in Fig.~\ref{fig:x3872_omega_E_plane}. Its isospin-breaking measure first becomes slightly larger than unity and then decreases toward zero as $\AIBX\to0$. In this same limit, the pole is pushed to infinity. Thus, while $W_{c1}$ is the pole most directly associated with the charged-channel bound state at $\AIBX=1$, it does not remain a nearby finite-energy pole when the imposed isospin-breaking component of the fixed $X(3872)$ pole vanishes.

The third accompanying pole, $W_{c1}^{\rm (s)}$, is the shadow of $W_{c1}$ at $\AIBX=1$. Its trajectory is exactly the complex conjugate of that of $W_{c1}$ when $\AIBX>0$. Together, $X^{\rm (s)}$, $W_{c1}$, and $W_{c1}^{\rm (s)}$ show how the pole content of the coupled-channel system evolves continuously between different decoupling limits. The motion is not a set of independent pole displacements. Rather, the poles exchange their roles as bound-state poles, shadow poles, and resonance-like poles as the relative neutral--charged composition of the fixed $X(3872)$ pole is varied.

Continuing the evolution from $\AIBX=0$ to $\AIBX=-1$ reverses the neutral--charged hierarchy of the endpoint. At $\AIBX=-1$, the two channels again decouple, but now the fixed $X(3872)$ pole couples only to the charged channel. The pole that was the shadow of the neutral $X(3872)$ at $\AIBX=1$ becomes the neutral-channel bound state at $\AIBX=-1$, with $A_{\rm IB}=1$.
Meanwhile, the charged-channel bound state and its shadow at $\AIBX=1$ become the shadow partners of the two bound-state poles at $\AIBX=-1$. In this limit, the other three poles lie on the real axes of different Riemann sheets below the neutral threshold. The full trajectory therefore interpolates between two opposite decoupled limits: one in which the fixed pole is purely neutral, and the other in which it is purely charged.

Two lessons for molecular systems can be drawn from this evolution. First, the companion pole is not an optional addition to the near-threshold dynamics. It is tied to the same isovector interaction that enhances the isospin-breaking component of the $X(3872)$. Second, depending on the interaction strength, the relevant pole may appear as a bound state, a virtual state, or a resonance-like pole, as reflected in the value of $\AIBX$.

The phenomenological implication is direct. For a near-threshold bound state such as the $X(3872)$, a larger isospin-breaking component implies a stronger interaction in the partner isospin channel. As a consequence, the corresponding companion pole becomes more relevant for physical line shapes. This mechanism is not specific to the $X(3872)$. The $T_{cc}(3875)^+$ observed by LHCb~\cite{LHCb:2021vvq,LHCb:2021auc} lies close to the $DD^*$ thresholds and is commonly interpreted as an isoscalar bound state with $\epsilon\simeq0.3$~\cite{Du:2021zzh}. Although this value is not as small as in the $X(3872)$ case, visible isospin-breaking effects can still arise if the isovector interaction is sizable. A similar situation occurs for the $P_c(4457)$~\cite{LHCb:2019kea}, which lies close to the $\Sigma_c\bar D^*$ threshold and can be interpreted as an $I=1/2$ bound state with $\epsilon\simeq0.6$~\cite{Du:2019pij,Liu:2019tjn}. In this case, a search for the isospin-breaking decay $P_c(4457)\to J/\psi\Delta^+$ would provide a direct probe of the isospin structure of the $P_c(4457)$ couplings. A sizable signal would indicate a strong interaction in the $I=3/2$ channel and would therefore support the existence of a nearby $I=3/2$ companion pole, whereas a stringent upper limit would disfavor such a scenario.\footnote{Sizable isospin-breaking decays $P_c(4457)\to J/\psi\Delta^+$ were predicted in Ref.~\cite{Guo:2019fdo} by considering only the threshold splitting between $\bar D^{*0}\Sigma_c^{+}$ and $D^{*-}\Sigma_c^{++}$. We emphasize, however, that such an estimate is scale dependent. From the present analysis, this decay can become sizable only if $\Sigma_c\bar D^*$ has a sizable $I=3/2$ interaction.} Conversely, the observation of such companion poles would imply sizable isospin breaking in the corresponding near-threshold states.

\subsection{Fate of poles in the isospin limit}

To identify which poles carry the isospin quantum numbers in the
symmetric limit, we reduce the threshold splitting $\Delta$ from its
physical value, $\Delta_{\rm phys}=8.23$~MeV, to zero while keeping the $S$-wave scattering lengths
$a_0^{I}$ ($I=0,1$), or equivalently the renormalized interaction strengths $C_I^{\rm R}$, fixed. The couplings are
chosen such that, at $\Delta=\Delta_{\rm phys}$, the $X(3872)$ has a
binding energy $E_b=160$~keV and $\AIBX=0.26$. This procedure removes
the kinematic source of isospin breaking without changing the
underlying interactions.

\begin{figure}[t]
 \centering
 \includegraphics[width=\linewidth]
 {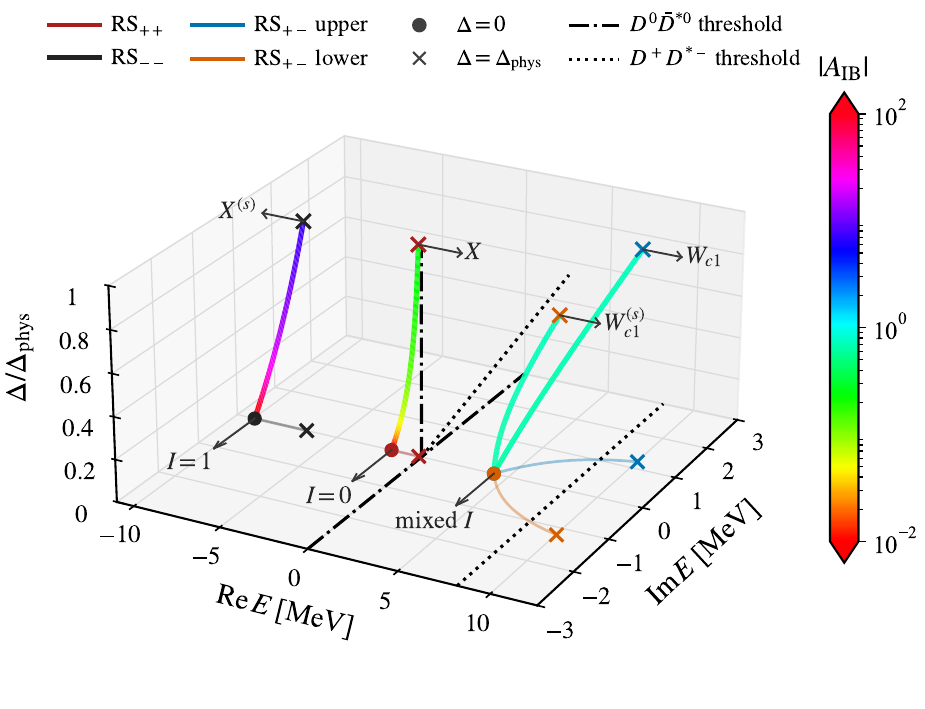}
 \caption{
 Pole trajectories as the threshold splitting $\Delta$ is reduced
 from its physical value to zero.
 Crosses and dots mark the poles at $\Delta=\Delta_{\rm phys}$ and
 $\Delta=0$, respectively. The interaction parameters are fixed at
 the values that produce an $X(3872)$ binding energy
 $E_b=160$~keV and $\AIBX=0.26$ at the physical threshold splitting.
 The vertical coordinate gives $\Delta/\Delta_{\rm phys}$, while
 the color along each three-dimensional trajectory represents the
 corresponding $|\AIB|$. The trajectories are projected onto the
 $\Delta=0$ plane, with the projection colors distinguishing the
 Riemann sheets. The dash-dotted and dotted black straight lines denote the neutral
 and charged thresholds, respectively.
 }
 \label{fig:delta_flow}
\end{figure}

Figure~\ref{fig:delta_flow} shows that the $X(3872)$ pole on
RS$_{++}$ remains on the real axis and moves from
$E_X=-0.16~\mathrm{MeV}$ at the physical point to
\begin{equation}
E_X\to-1.75~\mathrm{MeV},
\qquad
\AIBX\to0 ,
\end{equation}
as $\Delta\to0$, where the energy, as before, is measured relative to the neutral
threshold. The endpoint is therefore a pure $I=0$ bound state. The
finite isovector coupling of the physical $X(3872)$ disappears when the
two thresholds become degenerate, confirming that $\Delta$ supplies
the seed of its isospin breaking.

The fate of the putative $I=1$ companion is less intuitive. At the
physical point, the conjugate poles $W_{c1}$ and $W_{c1}^{\rm (s)}$ lie close to the
charged threshold on RS$_{+-}$,
\begin{equation}
E_{W_{c1},W_{c1}^{(s)}}=(10.03\pm1.20\,i)~\mathrm{MeV}.
\end{equation}
When $\Delta$ is reduced, these two poles move toward one another and
coalesce at
\begin{equation}
E_{W_{c1}^{\phantom{(s)}}}=E_{W_{c1}^{\rm (s)}}=4.17~\mathrm{MeV},
\quad
\AIB^{W_{c1},W_{c1}^{(s)}}=\pm0.65\,i .
\end{equation}
They therefore do not become pure $I=1$ poles, despite being the poles
closest to the physical region at $\Delta=\Delta_{\rm phys}$.

This behavior follows from the topology of the Riemann surface. For
nondegenerate thresholds, the neutral and charged cuts can be crossed
independently with different threshold branch points, giving four sheets. When the thresholds coincide,
however, an isospin-preserving analytic continuation must cross the
two degenerate cuts symmetrically. In the present notation, RS$_{++}$
and RS$_{--}$ are the symmetry-preserving sheets, whereas continuing
only one channel to RS$_{+-}$ or RS$_{-+}$ assigns different analytic
branches to two otherwise degenerate channels. The analytic
continuation itself then mixes $I=0$ and $I=1$, even though the
physical-axis amplitude has recovered isospin symmetry. Consequently,
the coalescing $W_{c1}$ and $W_{c1}^{\rm (s)}$ poles on the mixed sheet should not be
identified as additional isospin eigenstates of the symmetric theory.

Instead, the pure $I=1$ pole is reached along the fully continued
RS$_{--}$ trajectory. The pole denoted by $X^{\rm (s)}$ moves from
\begin{equation}
E_{X^{\rm (s)}}=-6.79~\mathrm{MeV}
\end{equation}
at the physical point to
\begin{equation}
E_{X^{\rm (s)}}\to-9.94~\mathrm{MeV},
\qquad
\AIB^{X^{\rm (s)}}\to\infty
\end{equation}
in the isospin limit. The
divergence of $\AIB$ means that at the endpoint the pole is purely $I=1$.

The pole continuously connected to the $I=1$ eigenstate is thus not
the nearby mixed-sheet pole at the physical point, but the more distant
fully continued shadow pole. This is the two-channel analogue of the interchange of roles between leading and shadow poles found when several thresholds merge in flavor
symmetry limits~\cite{Pelaez:2026fmu}, illustrating the general phenomenon that the pole which most
strongly affects the physical region away from the symmetric limit
need not be the pole that carries the corresponding symmetry quantum
numbers in that limit. Conversely, symmetry eigenstates that are well defined in the symmetric limit may evolve, once symmetry breaking is restored to its physical value, into poles located far from the physical region~\cite{Jido:2003cb,Guo:2023wes}, and therefore are less relevant for observables. Pole identity across a symmetry-restoring limit cannot be inferred
from continuity of its energy trajectory alone. One must also track
how the pole's Riemann sheet is connected to the physical region and
whether that analytic continuation preserves the restored symmetry~\cite{Pelaez:2026fmu}.

\section{Conclusion}\label{sec:summary}

We identify the dynamical conditions for amplified isospin breaking in a minimal near-threshold coupled-channel system. Large isospin breaking requires both threshold splitting and a sizable interaction in the companion isospin channel. A large 
$\AIB$ implies that the inverse amplitude in the companion isospin channel is driven to be small near the physical pole, which in turn entails the presence of a nearby companion pole, and the identification of such a companion state implies enhanced isospin-violating decays. This mechanism is directly testable through searches for partner poles and isospin-breaking decays in systems such as the $X(3872)$, the $T_{cc}$, and the $P_c$ states.

Within the system considered here, we also confirmed the general phenomenon of the interchange of shadow-pole role in coupled channel systems reported in a recent paper~\cite{Pelaez:2026fmu}. When nearby thresholds merge in a symmetry-restoring limit, the pole that dominates the line shape in the physical case need not evolve into the corresponding symmetry eigenstate. The
latter may instead be continuously connected to a more distant shadow
pole. 

A final remark is in order. If the production mechanism involves infrared singularities other than the two-body thresholds discussed here, such as triangle singularities, there can be further enhancement of isospin breaking. Such effects have been discussed in Ref.~\cite{Wu:2011yx} for the $\eta(1405/1475)\to 3\pi$ decay and are irrelevant here.

\begin{acknowledgments}
This work is supported in part by
Deutsche Forschungsgemeinschaft (DFG) under Grant No. 525056915
and under Germany's Excellence Strategy -- EXC 3107 -- Project-ID~533766364; by the National Natural Science Foundation of China (NSFC) under Grants No. 12125507, No. 12361141819, and No. 12447101; by the National Key R\&D Program of China under Grant No. 2023YFA1606703; and by the Chinese Academy of Sciences (CAS) under Grant No.~YSBR-101. In addition, U.-G.M. and C.H. thank the CAS President's International Fellowship Initiative (PIFI) under Grant Nos. 2025PD0022 and 2025PD0087, respectively, for partial support.
\end{acknowledgments}

\begin{appendix}

\section{Flavor structure of the short-distance $D\bar D^*\to J/\psi\mathcal V$ transitions}
\label{app:short-distance-flavor}

As discussed in the main text, the $\rho$ and $\omega$ are related through the light-flavor U(2) symmetry. The relation bears corrections from isospin breaking within triplet and from the relation between triplet and singlet. In this appendix, we discuss in more detail the flavor structure of the short-distance transitions $
D\bar D^*\to J/\psi\mathcal V,
$
and clarify the approximations entering the relation between the experimentally extracted coupling ratio $R_X$ and the isospin-breaking measure $\AIB$ determined from pole parameters.

The corresponding short-distance transition amplitudes are denoted by
$
u_{a\mathcal V}$ with $a=n,c$ and $
\mathcal V=\rho,\omega.
$
To make the light-flavor structure explicit, we introduce
\begin{equation}
P=(D^0, D^+),
\qquad
\bar P^*=(
\bar D^{*0},
D^{*-})^{\rm T},
\end{equation}
and the nonstrange light-vector-meson matrix
\begin{equation}
\mathbf{V}=
\frac{1}{\sqrt{2}}
\begin{pmatrix}
\omega+\rho^0 & \sqrt{2}\rho^+\\
\sqrt{2}\rho^- & \omega-\rho^0
\end{pmatrix}.
\label{eq:app-vector-matrix}
\end{equation}
The most general leading (nonderivative, without an insertion of the quark mass or charge matrix) interaction that satisfies the U(2) flavor symmetry contains two independent flavor structures (see Ref.~\cite{Meissner:2000bc} for a similar construction for the coupling between $J/\psi$, light vector mesons and a light scalar source),
\begin{equation}
\mathcal L_{\rm SD}=
J/\psi\left(
a\left\langle P \mathbf V\bar P^*\right\rangle
+b\left\langle P\bar P^*\right\rangle
\left\langle \mathbf V\right\rangle
\right)
+\mathrm{H.c.},
\label{eq:app-SD-Lagrangian}
\end{equation}
where the brackets denote traces in light-flavor space and common Lorentz, spin, and momentum structures have been suppressed.

For the neutral light vector mesons, Eq.~\eqref{eq:app-SD-Lagrangian} gives
\begin{align}
\mathcal L_{\rm SD}^{(0)}=&
J/\psi\Bigg[
\frac{a}{\sqrt{2}}
\rho^0
\left(
D^0\bar D^{*0}-D^+D^{*-}
\right)
\nonumber\\
&+
\frac{a+2b}{\sqrt{2}}
\omega
\left(
D^0\bar D^{*0}+D^+D^{*-}
\right)
\Bigg]
+\mathrm{H.c.}
\label{eq:app-neutral-SD}
\end{align}
Thus, 
\begin{align}
u_{n\rho}=-u_{c\rho}=\frac{a}{\sqrt{2}},\quad 
u_{n\omega}=u_{c\omega}=\frac{a+2b}{\sqrt{2}}.
\label{eq:app-V-couplings}
\end{align}
Genuine short-distance isospin breaking arises from the light-quark mass difference and electromagnetic interactions and scales as in Eq.~\eqref{eq:deltaI}.
Accordingly, it is subleading, and beyond the isospin relation we obtain in Eq.~\eqref{eq:u-deltaI}. 

The two terms in Eq.~\eqref{eq:app-SD-Lagrangian} correspond to different light-quark-line topologies. Schematically,
\begin{equation}
D_i\sim c\bar q_i,
\qquad
\bar D^{*i}\sim q^i\bar c,
\qquad
V_i{}^j\sim q_i\bar q^{\,j},
\end{equation}
with $i,j$ the light-flavor indices.
The single-trace interaction has the flavor structure (we do not show the $J/\psi$ to simplify the notation) $
\bar D^{*i}V_i{}^jD_j,$
for which the light-quark flavor flow is connected among the two heavy-light mesons and the light vector meson. In contrast, the double-trace term, $
(\bar D^{*i}D_i)(V_j{}^j),$
contains an independent flavor-singlet contraction; such terms correspond to disconnected---when gluon lines are neglected---quark loops and violate the OZI rule.
In order to mediate the interaction, at least two gluons need to be exchanged between the two quark loops (two flavor traces). 
Therefore, in the large $N_c$ counting, each additional flavor trace leads to a suppression of $\mathcal{O}(N_c N_c^{-2})=\mathcal{O}(N_c^{-1})$, with $N_c$ and $N_c^{-2}$ from the additional quark loop and four quark-gluon vertices, respectively~\cite{tHooft:1973alw,Witten:1979kh}; see also Ref.~\cite{Manohar:1998xv} for a pedagogical discussion on the relation between flavor traces and $N_c$ scaling.
Defining $\delta_{\rm OZI}\sim{b}/{a}$
and treating the double-trace interaction as a subleading correction, one obtains Eq.~\eqref{eq:OZI-relation} from Eq.~\eqref{eq:app-V-couplings}.

Near the $X(3872)$ pole, let $g_n$ and $g_c$ denote its couplings in the neutral and charged $D\bar D^*$ channels. The short-distance couplings to the hidden-charm vector final states are then
\begin{align}
g_{XJ/\psi\mathcal V}\propto g_nu_{n\mathcal V}+g_cu_{c\mathcal V}.
\end{align}
Using Eqs.~(\ref{eq:u-deltaI},\ref{eq:OZI-relation}), we obtain the ratio as in Eq.~\eqref{eq:RX-AIB}.
\end{appendix}

\bibliography{ref}

@article{Daub:2015xja,
    author = "Daub, J. T. and Hanhart, C. and Kubis, B.",
    title = "{A model-independent analysis of final-state interactions in $ {\overline{B}}_{d/s}^0\to J/\psi \pi \pi $}",
    eprint = "1508.06841",
    archivePrefix = "arXiv",
    primaryClass = "hep-ph",
    doi = "10.1007/JHEP02(2016)009",
    journal = "JHEP",
    volume = "02",
    pages = "009",
    year = "2016"
}

@article{Heuser:2025mnk,
    author = "Heuser, L. A. and Reyes-Torrecilla, A. and Hanhart, C. and Kubis, B. and Magalh{\~a}es, P. C. and Mannel, T. and Pel{\'a}ez, J. R.",
    title = "{Understanding Large Localized CP Violation in $B^{\pm}\to K^{\pm}\pi^+\pi^-$ Using Dispersive Methods}",
    eprint = "2508.10989",
    archivePrefix = "arXiv",
    primaryClass = "hep-ph",
    doi = "10.1103/l4rs-4znv",
    journal = "Phys. Rev. Lett.",
    volume = "136",
    number = "11",
    pages = "111901",
    year = "2026"
}

@article{Heuser:2026glv,
    author = "Heuser, L. A. and Reyes-Torrecilla, A. and Hanhart, C. and Huang, Y. and Kubis, B. and Magalh{\~a}es, P. C. and Mannel, T. and Pel{\'a}ez, J. R.",
    title = "{A dispersive method to study CP asymmetries in hadronic multi-body $B$ decays}",
    eprint = "2607.25764",
    archivePrefix = "arXiv",
    primaryClass = "hep-ph",
    journal = "",
    month = "7",
    year = "2026"
}

@article{ParticleDataGroup:2026aaa,
    author = "Takahashi, F. and others",
    collaboration = "Particle Data Group",
    title = "{Review of Particle Physics}",
    doi = "10.1142/S0217751X26300115",
    journal = "Int. J. Mod. Phys. A",
    volume = "41",
    pages = "2630011",
    year = "2026"
}

@article{Achasov:1979xc,
    author = "Achasov, N. N. and Devyanin, S. A. and Shestakov, G. N.",
    title = "{The $S^*$-$\delta^0$ Mixing as the Threshold Phenomenon}",
    reportNumber = "TF-106",
    doi = "10.1016/0370-2693(79)90488-X",
    journal = "Phys. Lett. B",
    volume = "88",
    pages = "367--371",
    year = "1979"
}

@article{Nomokonov:2002jb,
    author = "Nomokonov, V. P. and Sapozhnikov, M. G.",
    title = "{Experimental tests of the Okubo-Zweig-Iizuka rule in hadron interactions}",
    eprint = "hep-ph/0204259",
    archivePrefix = "arXiv",
    journal = "Phys. Part. Nucl.",
    volume = "34",
    pages = "94--123",
    year = "2003"
}

@article{Belle-II:2026xdx,
    author = "Abumusabh, M. and others",
    collaboration = "Belle-II, Belle",
    title = "{First evidence of $X(3872)\to\pi^0\chi_{c0}(1P)$ and search for $X(3915)\to\pi^0\chi_{c1}(1P)$}",
    eprint = "2606.24578",
    archivePrefix = "arXiv",
    primaryClass = "hep-ex",
    reportNumber = "Belle Preprint:2026-6, KEK Preprint: 2026-2",
    journal = "",
    month = "6",
    year = "2026"
}

@article{Maiani:2020zhr,
    author = "Maiani, Luciano and Polosa, Antonio D. and Riquer, Veronica",
    title = "{$X(3872)$ tetraquarks in $B$ and $B_s$ decays}",
    eprint = "2005.08764",
    archivePrefix = "arXiv",
    primaryClass = "hep-ph",
    doi = "10.1103/PhysRevD.102.034017",
    journal = "Phys. Rev. D",
    volume = "102",
    number = "3",
    pages = "034017",
    year = "2020"
}

@article{Maiani:2017kyi,
    author = "Maiani, L. and Polosa, A. D. and Riquer, V.",
    title = "{A Theory of X and Z Multiquark Resonances}",
    eprint = "1712.05296",
    archivePrefix = "arXiv",
    primaryClass = "hep-ph",
    doi = "10.1016/j.physletb.2018.01.039",
    journal = "Phys. Lett. B",
    volume = "778",
    pages = "247--251",
    year = "2018"
}

@article{Hanhart:2007bd,
    author = "Hanhart, Christoph and Kubis, Bastian and Pelaez, Jose R.",
    title = "{Investigation of $a_0$-$f_0$ mixing}",
    eprint = "0707.0262",
    archivePrefix = "arXiv",
    primaryClass = "hep-ph",
    reportNumber = "FZJ-IKP-TH-2007-21, HISKP-TH-07-18",
    doi = "10.1103/PhysRevD.76.074028",
    journal = "Phys. Rev. D",
    volume = "76",
    pages = "074028",
    year = "2007"
}

@article{Kato:1965iee,
    author = "Kato, Masaaki",
    title = "{Analytical properties of two-channel $S$-matrix}",
    doi = "10.1016/0003-4916(65)90235-6",
    journal = "Annals Phys.",
    volume = "31",
    number = "1",
    pages = "130--147",
    year = "1965"
}

@article{Eden:1964zz,
    author = "Eden, R. J. and Taylor, J. R.",
    title = "{Poles and Shadow Poles in the Many-Channel S Matrix}",
    doi = "10.1103/PhysRev.133.B1575",
    journal = "Phys. Rev.",
    volume = "133",
    pages = "B1575--B1580",
    year = "1964"
}

@article{Dong:2021juy,
    author = "Dong, Xiang-Kun and Guo, Feng-Kun and Zou, Bing-Song",
    title = "{A survey of heavy-antiheavy hadronic molecules}",
    eprint = "2101.01021",
    archivePrefix = "arXiv",
    primaryClass = "hep-ph",
    doi = "10.13725/j.cnki.pip.2021.02.001",
    journal = "Progr. Phys.",
    volume = "41",
    pages = "65--93",
    year = "2021"
}

@article{Baru:2021ldu,
    author = "Baru, Vadim and Dong, Xiang-Kun and Du, Meng-Lin and Filin, Arseniy and Guo, Feng-Kun and Hanhart, Christoph and Nefediev, Alexey and Nieves, Juan and Wang, Qian",
    title = "{Effective range expansion for narrow near-threshold resonances}",
    eprint = "2110.07484",
    archivePrefix = "arXiv",
    primaryClass = "hep-ph",
    doi = "10.1016/j.physletb.2022.137290",
    journal = "Phys. Lett. B",
    volume = "833",
    pages = "137290",
    year = "2022"
}

@article{Meissner:2000bc,
    author = "Mei\ss{}ner, Ulf-G. and Oller, J. A.",
    title = "{$J/\psi\to\phi\pi\pi(K\bar K)$ decays, chiral dynamics and OZI violation}",
    eprint = "hep-ph/0005253",
    archivePrefix = "arXiv",
    reportNumber = "FZJ-IKP-TH-2000-12",
    doi = "10.1016/S0375-9474(00)00367-5",
    journal = "Nucl. Phys. A",
    volume = "679",
    pages = "671--697",
    year = "2001"
}

@article{Iizuka:1966fk,
    author = "Iizuka, Jugoro",
    title = "{Systematics and phenomenology of meson family}",
    doi = "10.1143/PTPS.37.21",
    journal = "Prog. Theor. Phys. Suppl.",
    volume = "37",
    pages = "21--34",
    year = "1966"
}

@article{Okubo:1963fa,
    author = "Okubo, S.",
    title = "{$\phi$ meson and unitary symmetry model}",
    doi = "10.1016/S0375-9601(63)92548-9",
    journal = "Phys. Lett.",
    volume = "5",
    pages = "165--168",
    year = "1963"
}

@article{Witten:1979kh,
    author = "Witten, Edward",
    title = "{Baryons in the $1/N$ Expansion}",
    reportNumber = "HUTP-79-A007",
    doi = "10.1016/0550-3213(79)90232-3",
    journal = "Nucl. Phys. B",
    volume = "160",
    pages = "57--115",
    year = "1979"
}

@article{tHooft:1973alw,
    author = "'t Hooft, Gerard",
    editor = "Taylor, J. C.",
    title = "{A Planar Diagram Theory for Strong Interactions}",
    reportNumber = "CERN-TH-1786",
    doi = "10.1016/0550-3213(74)90154-0",
    journal = "Nucl. Phys. B",
    volume = "72",
    pages = "461",
    year = "1974"
}

@article{Guo:2023wes,
    author = "Guo, Feng-Kun and Kamiya, Yuki and Mai, Maxim and Mei{\ss}ner, Ulf-G.",
    title = "{New insights into the nature of the {\ensuremath{\Lambda}}(1380) and {\ensuremath{\Lambda}}(1405) resonances away from the SU(3) limit}",
    eprint = "2308.07658",
    archivePrefix = "arXiv",
    primaryClass = "hep-ph",
    doi = "10.1016/j.physletb.2023.138264",
    journal = "Phys. Lett. B",
    volume = "846",
    pages = "138264",
    year = "2023"
}

@article{Jido:2003cb,
    author = "Jido, D. and Oller, J. A. and Oset, E. and Ramos, A. and Mei{\ss}ner, U.-G.",
    title = "{Chiral dynamics of the two $\Lambda(1405)$ states}",
    eprint = "nucl-th/0303062",
    archivePrefix = "arXiv",
    doi = "10.1016/S0375-9474(03)01598-7",
    journal = "Nucl. Phys. A",
    volume = "725",
    pages = "181--200",
    year = "2003"
}

@article{Sibirtsev:2005nq,
    author = "Sibirtsev, Alexander and Mei\ss{}ner, Ulf-G. and Thomas, Anthony William",
    title = "{OZI rule violation in photoproduction}",
    eprint = "hep-ph/0503276",
    archivePrefix = "arXiv",
    reportNumber = "FZJ-IKP-TH-2005-10, HISKP-TH-05-07, JLAB-THY-05-309",
    doi = "10.1103/PhysRevD.71.094011",
    journal = "Phys. Rev. D",
    volume = "71",
    pages = "094011",
    year = "2005"
}

@article{Wu:2021udi,
    author = "Wu, Qi and Chen, Dian-Yong and Matsuki, Takayuki",
    title = "{A phenomenological analysis on isospin-violating decay of $X(3872)$}",
    eprint = "2102.08637",
    archivePrefix = "arXiv",
    primaryClass = "hep-ph",
    doi = "10.1140/epjc/s10052-021-08984-2",
    journal = "Eur. Phys. J. C",
    volume = "81",
    number = "2",
    pages = "193",
    year = "2021"
}

@article{Takeuchi:2014rsa,
    author = "Takeuchi, Sachiko and Shimizu, Kiyotaka and Takizawa, Makoto",
    title = "{On the origin of the narrow peak and the isospin symmetry breaking of the $X$(3872)}",
    eprint = "1408.0973",
    archivePrefix = "arXiv",
    primaryClass = "hep-ph",
    reportNumber = "RIKEN-QHP-165",
    doi = "10.1093/ptep/ptv104",
    journal = "PTEP",
    volume = "2014",
    number = "12",
    pages = "123D01",
    year = "2014",
    note = "[Erratum: PTEP 2015, 079203 (2015)]"
}

@article{Li:2012cs,
    author = "Li, Ning and Zhu, Shi-Lin",
    title = "{Isospin breaking, Coupled-channel effects and Diagnosis of X(3872)}",
    eprint = "1207.3954",
    archivePrefix = "arXiv",
    primaryClass = "hep-ph",
    doi = "10.1103/PhysRevD.86.074022",
    journal = "Phys. Rev. D",
    volume = "86",
    pages = "074022",
    year = "2012"
}

@article{Hidalgo-Duque:2012rqv,
    author = "Hidalgo-Duque, C. and Nieves, J. and Valderrama, M. Pavon",
    title = "{Light flavor and heavy quark spin symmetry in heavy meson molecules}",
    eprint = "1210.5431",
    archivePrefix = "arXiv",
    primaryClass = "hep-ph",
    doi = "10.1103/PhysRevD.87.076006",
    journal = "Phys. Rev. D",
    volume = "87",
    number = "7",
    pages = "076006",
    year = "2013"
}

@article{Zhang:2024fxy,
    author = "Zhang, Zhen-Hua and Ji, Teng and Dong, Xiang-Kun and Guo, Feng-Kun and Hanhart, Christoph and Mei{\ss}ner, Ulf-G. and Rusetsky, Akaki",
    title = "{Predicting isovector charmonium-like states from X(3872) properties}",
    eprint = "2404.11215",
    archivePrefix = "arXiv",
    primaryClass = "hep-ph",
    doi = "10.1007/JHEP08(2024)130",
    journal = "JHEP",
    volume = "08",
    pages = "130",
    year = "2024"
}

@article{Meng:2021kmi,
    author = "Meng, Lu and Wang, Guang-Juan and Wang, Bo and Zhu, Shi-Lin",
    title = "{Revisit the isospin violating decays of X(3872)}",
    eprint = "2109.01333",
    archivePrefix = "arXiv",
    primaryClass = "hep-ph",
    doi = "10.1103/PhysRevD.104.094003",
    journal = "Phys. Rev. D",
    volume = "104",
    number = "9",
    pages = "094003",
    year = "2021"
}

@article{LHCb:2020xds,
    author = "Aaij, R. and others",
    collaboration = "LHCb",
    title = "{Study of the lineshape of the $\chi_{c1}(3872)$ state}",
    eprint = "2005.13419",
    archivePrefix = "arXiv",
    primaryClass = "hep-ex",
    reportNumber = "CERN-EP-2020-086, LHCb-PAPER-2020-008",
    doi = "10.1103/PhysRevD.102.092005",
    journal = "Phys. Rev. D",
    volume = "102",
    number = "9",
    pages = "092005",
    year = "2020"
}

@article{BESIII:2023hml,
    author = "Ablikim, Medina and others",
    collaboration = "BESIII",
    title = "{Coupled-Channel Analysis of the $\chi_{c1}(3872)$ Line Shape with BESIII Data}",
    eprint = "2309.01502",
    archivePrefix = "arXiv",
    primaryClass = "hep-ex",
    doi = "10.1103/PhysRevLett.132.151903",
    journal = "Phys. Rev. Lett.",
    volume = "132",
    number = "15",
    pages = "151903",
    year = "2024"
}

@article{Sone:2024nfj,
    author = "Sone, Katsuyoshi and Hyodo, Tetsuo",
    title = "{General amplitude of near-threshold hadron scattering for exotic hadrons}",
    eprint = "2405.08436",
    archivePrefix = "arXiv",
    primaryClass = "hep-ph",
    doi = "10.1103/qp1j-slqg",
    journal = "Phys. Rev. C",
    volume = "112",
    number = "6",
    pages = "065207",
    year = "2025"
}

@article{Liu:2019tjn,
    author = "Liu, Ming-Zhu and Pan, Ya-Wen and Peng, Fang-Zheng and S{\'a}nchez S{\'a}nchez, Mario and Geng, Li-Sheng and Hosaka, Atsushi and Pavon Valderrama, Manuel",
    title = "{Emergence of a complete heavy-quark spin symmetry multiplet: seven molecular pentaquarks in light of the latest LHCb analysis}",
    eprint = "1903.11560",
    archivePrefix = "arXiv",
    primaryClass = "hep-ph",
    doi = "10.1103/PhysRevLett.122.242001",
    journal = "Phys. Rev. Lett.",
    volume = "122",
    number = "24",
    pages = "242001",
    year = "2019"
}

@article{Du:2019pij,
    author = "Du, Meng-Lin and Baru, Vadim and Guo, Feng-Kun and Hanhart, Christoph and Mei{\ss}ner, Ulf-G and Oller, Jos{\'e} A. and Wang, Qian",
    title = "{Interpretation of the LHCb $P_c$ States as Hadronic Molecules and Hints of a Narrow $P_c(4380)$}",
    eprint = "1910.11846",
    archivePrefix = "arXiv",
    primaryClass = "hep-ph",
    doi = "10.1103/PhysRevLett.124.072001",
    journal = "Phys. Rev. Lett.",
    volume = "124",
    number = "7",
    pages = "072001",
    year = "2020"
}

@article{Gamermann:2009uq,
    author = "Gamermann, D. and Nieves, J. and Oset, E. and Ruiz Arriola, E.",
    title = "{Couplings in coupled channels versus wave functions: application to the $X(3872)$ resonance}",
    eprint = "0911.4407",
    archivePrefix = "arXiv",
    primaryClass = "hep-ph",
    doi = "10.1103/PhysRevD.81.014029",
    journal = "Phys. Rev. D",
    volume = "81",
    pages = "014029",
    year = "2010"
}

@article{Belle:2011vlx,
    author = "Choi, S.-K. and others",
    collaboration = "Belle",
    title = "{Bounds on the width, mass difference and other properties of $X(3872) \to \pi^+ \pi^- J/\psi$ decays}",
    eprint = "1107.0163",
    archivePrefix = "arXiv",
    primaryClass = "hep-ex",
    doi = "10.1103/PhysRevD.84.052004",
    journal = "Phys. Rev. D",
    volume = "84",
    pages = "052004",
    year = "2011"
}

@article{BaBar:2004cah,
    author = "Aubert, Bernard and others",
    collaboration = "BaBar",
    title = "{Search for a charged partner of the $X(3872) $in the $B$ meson decay $B \to X^- K$, $X^- \to J/\psi \pi^- \pi^0$}",
    eprint = "hep-ex/0412051",
    archivePrefix = "arXiv",
    reportNumber = "SLAC-PUB-10903, BABAR-PUB-04-043",
    doi = "10.1103/PhysRevD.71.031501",
    journal = "Phys. Rev. D",
    volume = "71",
    pages = "031501",
    year = "2005"
}

@article{LHCb:2022jez,
    author = "Aaij, Roel and others",
    collaboration = "LHCb",
    title = "{Observation of sizeable $\omega$ contribution to $\chi_{c1}(3872) \to \pi^+\pi^- J/\psi$ decays}",
    eprint = "2204.12597",
    archivePrefix = "arXiv",
    primaryClass = "hep-ex",
    reportNumber = "LHCb-PAPER-2021-045, CERN-EP-2022-049",
    doi = "10.1103/PhysRevD.108.L011103",
    journal = "Phys. Rev. D",
    volume = "108",
    number = "1",
    pages = "L011103",
    year = "2023"
}

@article{BESIII:2019esk,
    author = "Ablikim, M. and others",
    collaboration = "BESIII",
    title = "{Observation of the decay $X(3872) \to \pi^0 \chi_{c1}(1P)$}",
    eprint = "1901.03992",
    archivePrefix = "arXiv",
    primaryClass = "hep-ex",
    doi = "10.1103/PhysRevLett.122.202001",
    journal = "Phys. Rev. Lett.",
    volume = "122",
    number = "20",
    pages = "202001",
    year = "2019"
}

@article{Suzuki:2005ha,
    author = "Suzuki, Mahiko",
    title = "{The $X(3872)$ boson: Molecule or charmonium}",
    eprint = "hep-ph/0508258",
    archivePrefix = "arXiv",
    reportNumber = "LBL-58724",
    doi = "10.1103/PhysRevD.72.114013",
    journal = "Phys. Rev. D",
    volume = "72",
    pages = "114013",
    year = "2005"
}

@article{LHCb:2021auc,
    author = "Aaij, Roel and others",
    collaboration = "LHCb",
    title = "{Study of the doubly charmed tetraquark $T_{cc}^{+}$}",
    eprint = "2109.01056",
    archivePrefix = "arXiv",
    primaryClass = "hep-ex",
    reportNumber = "CERN-EP-2021-169, LHCb-PAPER-2021-032",
    doi = "10.1038/s41467-022-30206-w",
    journal = "Nature Commun.",
    volume = "13",
    number = "1",
    pages = "3351",
    year = "2022"
}

@article{LHCb:2019kea,
    author = "Aaij, Roel and others",
    collaboration = "LHCb",
    title = "{Observation of a narrow pentaquark state, $P_c(4312)^+$, and of two-peak structure of the $P_c(4450)^+$}",
    eprint = "1904.03947",
    archivePrefix = "arXiv",
    primaryClass = "hep-ex",
    reportNumber = "LHCb-PAPER-2019-014 CERN-EP-2019-058",
    doi = "10.1103/PhysRevLett.122.222001",
    journal = "Phys. Rev. Lett.",
    volume = "122",
    number = "22",
    pages = "222001",
    year = "2019"
}

@article{Guo:2019fdo,
    author = "Guo, Feng-Kun and Jing, Hao-Jie and Mei{\ss}ner, Ulf-G and Sakai, Shuntaro",
    title = "{Isospin breaking decays as a diagnosis of the hadronic molecular structure of the $P_c(4457)$}",
    eprint = "1903.11503",
    archivePrefix = "arXiv",
    primaryClass = "hep-ph",
    doi = "10.1103/PhysRevD.99.091501",
    journal = "Phys. Rev. D",
    volume = "99",
    number = "9",
    pages = "091501",
    year = "2019"
}

@article{Du:2021zzh,
    author = "Du, Meng-Lin and Baru, Vadim and Dong, Xiang-Kun and Filin, Arseniy and Guo, Feng-Kun and Hanhart, Christoph and Nefediev, Alexey and Nieves, Juan and Wang, Qian",
    title = "{Coupled-channel approach to $T_{cc}^+$ including three-body effects}",
    eprint = "2110.13765",
    archivePrefix = "arXiv",
    primaryClass = "hep-ph",
    doi = "10.1103/PhysRevD.105.014024",
    journal = "Phys. Rev. D",
    volume = "105",
    number = "1",
    pages = "014024",
    year = "2022"
}

@article{LHCb:2021vvq,
    author = "Aaij, Roel and others",
    collaboration = "LHCb",
    title = "{Observation of an exotic narrow doubly charmed tetraquark}",
    eprint = "2109.01038",
    archivePrefix = "arXiv",
    primaryClass = "hep-ex",
    reportNumber = "CERN-EP-2021-165, LHCb-PAPER-2021-031",
    doi = "10.1038/s41567-022-01614-y",
    journal = "Nature Phys.",
    volume = "18",
    number = "7",
    pages = "751--754",
    year = "2022"
}

@article{Tornqvist:2004qy,
    author = "T{\"o}rnqvist, Nils A.",
    title = "{Isospin breaking of the narrow charmonium state of Belle at 3872~MeV as a deuson}",
    eprint = "hep-ph/0402237",
    archivePrefix = "arXiv",
    doi = "10.1016/j.physletb.2004.03.077",
    journal = "Phys. Lett. B",
    volume = "590",
    pages = "209--215",
    year = "2004"
}

@article{Swanson:2003tb,
    author = "Swanson, Eric S.",
    title = "{Short range structure in the $X(3872)$}",
    eprint = "hep-ph/0311229",
    archivePrefix = "arXiv",
    reportNumber = "JLAB-THY-03-227",
    doi = "10.1016/j.physletb.2004.03.033",
    journal = "Phys. Lett. B",
    volume = "588",
    pages = "189--195",
    year = "2004"
}

@article{CDF:2003cab,
    author = "Acosta, D. and others",
    collaboration = "CDF",
    title = "{Observation of the narrow state $X(3872) \to J/\psi \pi^+ \pi^-$ in $\bar{p}p$ collisions at $\sqrt{s} = 1.96$ TeV}",
    eprint = "hep-ex/0312021",
    archivePrefix = "arXiv",
    reportNumber = "FERMILAB-PUB-03-393-E",
    doi = "10.1103/PhysRevLett.93.072001",
    journal = "Phys. Rev. Lett.",
    volume = "93",
    pages = "072001",
    year = "2004"
}

@article{Maiani:2004vq,
    author = "Maiani, L. and Piccinini, F. and Polosa, A. D. and Riquer, V.",
    title = "{Diquark-antidiquarks with hidden or open charm and the nature of $X(3872)$}",
    eprint = "hep-ph/0412098",
    archivePrefix = "arXiv",
    reportNumber = "ROMA1-1396-2004, FNT-T-2004-20, BA-TH-502-04, CERN-PH-TH-2004-239",
    doi = "10.1103/PhysRevD.71.014028",
    journal = "Phys. Rev. D",
    volume = "71",
    pages = "014028",
    year = "2005"
}

@article{D0:2004zmu,
    author = "Abazov, V. M. and others",
    collaboration = "D0",
    title = "{Observation and properties of the $X(3872)$ decaying to $J/\psi \pi^+ \pi^-$ in $p\bar{p}$ collisions at $\sqrt{s} = 1.96$ TeV}",
    eprint = "hep-ex/0405004",
    archivePrefix = "arXiv",
    reportNumber = "FERMILAB-PUB-04-061-E",
    doi = "10.1103/PhysRevLett.93.162002",
    journal = "Phys. Rev. Lett.",
    volume = "93",
    pages = "162002",
    year = "2004"
}

@article{CDF:2005cfq,
    author = "Abulencia, A. and others",
    collaboration = "CDF",
    title = "{Measurement of the dipion mass spectrum in $X(3872) \to J/\psi \pi^+ \pi^-$ decays.}",
    eprint = "hep-ex/0512074",
    archivePrefix = "arXiv",
    reportNumber = "FERMILAB-PUB-05-535-E",
    doi = "10.1103/PhysRevLett.96.102002",
    journal = "Phys. Rev. Lett.",
    volume = "96",
    pages = "102002",
    year = "2006"
}

@article{Gamermann:2009fv,
    author = "Gamermann, Daniel and Oset, Eulogio",
    title = "{Isospin breaking effects in the X(3872) resonance}",
    eprint = "0905.0402",
    archivePrefix = "arXiv",
    primaryClass = "hep-ph",
    doi = "10.1103/PhysRevD.80.014003",
    journal = "Phys. Rev. D",
    volume = "80",
    pages = "014003",
    year = "2009"
}

@article{Braaten:2003he,
    author = "Braaten, Eric and Kusunoki, Masaoki",
    title = "{Low-energy universality and the new charmonium resonance at 3870 MeV}",
    eprint = "hep-ph/0311147",
    archivePrefix = "arXiv",
    journal = "Phys. Rev. D",
    volume = "69",
    pages = "074005",
    year = "2004"
}

@article{Hanhart:2007yq,
    author = "Hanhart, C and Kalashnikova, Yu. S and Kudryavtsev, Alexander Evgenyevich and Nefediev, A. V",
    title = "{Reconciling the $X(3872)$ with the near-threshold enhancement in the $D^0 \bar{D}^{*0}$ final state}",
    eprint = "0704.0605",
    archivePrefix = "arXiv",
    primaryClass = "hep-ph",
    reportNumber = "FZJ-IKP-TH-2007-14",
    doi = "10.1103/PhysRevD.76.034007",
    journal = "Phys. Rev. D",
    volume = "76",
    pages = "034007",
    year = "2007"
}

@article{Guo:2017jvc,
    author = "Guo, Feng-Kun and Hanhart, Christoph and Mei{\ss}ner, Ulf-G. and Wang, Qian and Zhao, Qiang and Zou, Bing-Song",
    title = "{Hadronic molecules}",
    eprint = "1705.00141",
    archivePrefix = "arXiv",
    primaryClass = "hep-ph",
    doi = "10.1103/RevModPhys.90.015004",
    journal = "Rev. Mod. Phys.",
    volume = "90",
    number = "1",
    pages = "015004",
    year = "2018",
    note = "[Erratum: Rev.Mod.Phys. 94, 029901 (2022)]"
}

@inbook{Zweig:1964jf,
    author = "Zweig, G.",
    editor = "Lichtenberg, D. B. and Rosen, Simon Peter",
    title = "{An SU(3) model for strong interaction symmetry and its breaking. Version 2}",
    booktitle = "{Developments in the Quark Theory of Hadrons. Vol. 1. 1964--1978}",
    reportNumber = "CERN-TH-412, NP-14146, PRINT-64-170",
    pages = "22--101",
    month = "2",
    year = "1964"
}

@article{Dai:2026fkg,
    author = "Dai, Xinchen and Jia, Sen and Nefediev, Alexey and Nieves, Juan and Shen, Chengping and Zhang, Liming",
    title = "{Exotic hadrons associated with $b$-quark}",
    eprint = "2603.09315",
    archivePrefix = "arXiv",
    primaryClass = "hep-ph",
    doi = "10.1016/j.physrep.2026.06.002",
    journal = "Phys. Rept.",
    volume = "1191",
    pages = "1--62",
    year = "2026"
}

@article{Ji:2025hjw,
    author = "Ji, Teng and Dong, Xiang-Kun and Guo, Feng-Kun and Hanhart, Christoph and Mei\ss{}ner, Ulf-G.",
    title = "{Precise determination of the properties of $X(3872)$ and of its isovector partner $W_{c1}$}",
    doi = "10.1016/j.scib.2026.07.071",
    journal = "Sci. Bull.",
    eprint = "2502.04458",
    archivePrefix = "arXiv",
    primaryClass = "hep-ph",
    month = "2",
    year = "2025"
}

@article{Zhang:2024qkg,
    author = "Zhang, Zhen-Hua and Guo, Feng-Kun",
    title = "{Classification of coupled-channel near-threshold structures}",
    eprint = "2407.10620",
    archivePrefix = "arXiv",
    primaryClass = "hep-ph",
    doi = "10.1016/j.physletb.2025.139387",
    journal = "Phys. Lett. B",
    volume = "863",
    pages = "139387",
    year = "2025"
}

@article{Dong:2020hxe,
    author = "Dong, Xiang-Kun and Guo, Feng-Kun and Zou, Bing-Song",
    title = "{Explaining the Many Threshold Structures in the Heavy-Quark Hadron Spectrum}",
    eprint = "2011.14517",
    archivePrefix = "arXiv",
    primaryClass = "hep-ph",
    doi = "10.1103/PhysRevLett.126.152001",
    journal = "Phys. Rev. Lett.",
    volume = "126",
    number = "15",
    pages = "152001",
    year = "2021"
}

@article{Guo:2014iya,
    author = "Guo, Feng-Kun and Hanhart, Christoph and Wang, Qian and Zhao, Qiang",
    title = "{Could the near-threshold $XYZ$ states be simply kinematic effects?}",
    eprint = "1411.5584",
    archivePrefix = "arXiv",
    primaryClass = "hep-ph",
    doi = "10.1103/PhysRevD.91.051504",
    journal = "Phys. Rev. D",
    volume = "91",
    number = "5",
    pages = "051504",
    year = "2015"
}

@article{Guo:2019twa,
    author = "Guo, Feng-Kun and Liu, Xiao-Hai and Sakai, Shuntaro",
    title = "{Threshold cusps and triangle singularities in hadronic reactions}",
    eprint = "1912.07030",
    archivePrefix = "arXiv",
    primaryClass = "hep-ph",
    doi = "10.1016/j.ppnp.2020.103757",
    journal = "Prog. Part. Nucl. Phys.",
    volume = "112",
    pages = "103757",
    year = "2020"
}

@article{Ji:2022blw,
    author = "Ji, Teng and Dong, Xiang-Kun and Guo, Feng-Kun and Zou, Bing-Song",
    title = "{Prediction of a Narrow Exotic Hadronic State with Quantum Numbers $J^{PC}=0^{--}$}",
    eprint = "2205.10994",
    archivePrefix = "arXiv",
    primaryClass = "hep-ph",
    doi = "10.1103/PhysRevLett.129.102002",
    journal = "Phys. Rev. Lett.",
    volume = "129",
    number = "10",
    pages = "102002",
    year = "2022"
}

@article{Hosaka:2016pey,
    author = "Hosaka, Atsushi and Iijima, Toru and Miyabayashi, Kenkichi and Sakai, Yoshihide and Yasui, Shigehiro",
    title = "{Exotic hadrons with heavy flavors: $X$, $Y$, $Z$, and related states}",
    eprint = "1603.09229",
    archivePrefix = "arXiv",
    primaryClass = "hep-ph",
    reportNumber = "J-PARC-TH-0046",
    doi = "10.1093/ptep/ptw045",
    journal = "PTEP",
    volume = "2016",
    number = "6",
    pages = "062C01",
    year = "2016"
}

@article{Ali:2017jda,
    author = {Ali, Ahmed and Lange, Jens S{\"o}ren and Stone, Sheldon},
    title = "{Exotics: Heavy Pentaquarks and Tetraquarks}",
    eprint = "1706.00610",
    archivePrefix = "arXiv",
    primaryClass = "hep-ph",
    reportNumber = "DESY-17-071",
    doi = "10.1016/j.ppnp.2017.08.003",
    journal = "Prog. Part. Nucl. Phys.",
    volume = "97",
    pages = "123--198",
    year = "2017"
}

@article{Belle:2003nnu,
    author = "Choi, S. K. and others",
    collaboration = "Belle",
    title = "{Observation of a narrow charmonium-like state in exclusive $B^\pm \to K^\pm \pi^+ \pi^- J/\psi$ decays}",
    eprint = "hep-ex/0309032",
    archivePrefix = "arXiv",
    doi = "10.1103/PhysRevLett.91.262001",
    journal = "Phys. Rev. Lett.",
    volume = "91",
    pages = "262001",
    year = "2003"
}

@article{Olsen:2017bmm,
    author = "Olsen, Stephen Lars and Skwarnicki, Tomasz and Zieminska, Daria",
    title = "{Nonstandard heavy mesons and baryons: Experimental evidence}",
    eprint = "1708.04012",
    archivePrefix = "arXiv",
    primaryClass = "hep-ph",
    doi = "10.1103/RevModPhys.90.015003",
    journal = "Rev. Mod. Phys.",
    volume = "90",
    number = "1",
    pages = "015003",
    year = "2018"
}

@article{Esposito:2016noz,
    author = "Esposito, A. and Pilloni, A. and Polosa, A. D.",
    title = "{Multiquark Resonances}",
    eprint = "1611.07920",
    archivePrefix = "arXiv",
    primaryClass = "hep-ph",
    reportNumber = "JLAB-THY-16-2301",
    doi = "10.1016/j.physrep.2016.11.002",
    journal = "Phys. Rept.",
    volume = "668",
    pages = "1--97",
    year = "2017"
}

@article{Lebed:2016hpi,
    author = "Lebed, Richard F. and Mitchell, Ryan E. and Swanson, Eric S.",
    title = "{Heavy-Quark QCD Exotica}",
    eprint = "1610.04528",
    archivePrefix = "arXiv",
    primaryClass = "hep-ph",
    doi = "10.1016/j.ppnp.2016.11.003",
    journal = "Prog. Part. Nucl. Phys.",
    volume = "93",
    pages = "143--194",
    year = "2017"
}

@article{Chen:2022asf,
    author = "Chen, Hua-Xing and Chen, Wei and Liu, Xiang and Liu, Yan-Rui and Zhu, Shi-Lin",
    title = "{An updated review of the new hadron states}",
    eprint = "2204.02649",
    archivePrefix = "arXiv",
    primaryClass = "hep-ph",
    doi = "10.1088/1361-6633/aca3b6",
    journal = "Rept. Prog. Phys.",
    volume = "86",
    number = "2",
    pages = "026201",
    year = "2023"
}

@article{Belle:2008fma,
    author = "Aushev, T. and others",
    collaboration = "Belle",
    title = "{Study of the {$B\to X(3872)(D^{*0}\bar D^0)K$} decay}",
    eprint = "0810.0358",
    archivePrefix = "arXiv",
    primaryClass = "hep-ex",
    reportNumber = "BELLE-CONF-0832",
    doi = "10.1103/PhysRevD.81.031103",
    journal = "Phys. Rev. D",
    volume = "81",
    pages = "031103",
    year = "2010"
}

@article{Dias:2024zfh,
    author = "Dias, Jorgivan Morais and Ji, Teng and Dong, Xiang-Kun and Guo, Feng-Kun and Hanhart, Christoph and Mei\ss{}ner, Ulf-G. and Zhang, Yu and Zhang, Zhen-Hua",
    title = "{Dispersive analysis of the isospin breaking in the {$X(3872)\to J/\psi\pi^+\pi^-$} and {{$X(3872)\to J/\psi\pi^+\pi^0\pi^-$}} decays}",
    eprint = "2409.13245",
    archivePrefix = "arXiv",
    primaryClass = "hep-ph",
    doi = "10.1103/PhysRevD.111.014031",
    journal = "Phys. Rev. D",
    volume = "111",
    number = "1",
    pages = "014031",
    year = "2025"
}

@article{Wang:2022vjm,
    author = "Wang, Hao-Nan and Wang, Qian and Xie, Ju-Jun",
    title = "{Theoretical study on the contributions of $\omega$ meson 
    to the $X(3872)\to J/\psi\pi^+\pi^-$ and $J/\psi\pi^+\pi^-\pi^0$ decays}",
    eprint = "2206.14456",
    archivePrefix = "arXiv",
    primaryClass = "hep-ph",
    doi = "10.1103/PhysRevD.106.056022",
    journal = "Phys. Rev. D",
    volume = "106",
    number = "5",
    pages = "056022",
    year = "2022"
}

@article{Hanhart:2011tn,
    author = "Hanhart, C. and Kalashnikova, Yu. S. and Kudryavtsev, A. E. and Nefediev, A. V.",
    title = "{Remarks on the quantum numbers of $X(3872)$ from the invariant 
    mass distributions of the $\rho J/\psi$ and $\omega J/\psi$ final states}",
    eprint = "1111.6241",
    archivePrefix = "arXiv",
    primaryClass = "hep-ph",
    doi = "10.1103/PhysRevD.85.011501",
    journal = "Phys. Rev. D",
    volume = "85",
    pages = "011501",
    year = "2012"
    }

@article{Braaten:2005ai,
    author = "Braaten, Eric and Kusunoki, Masaoki",
    title = "{Decays of the $X(3872)$ into $J/\psi$ and light hadrons}",
    eprint = "hep-ph/0507163",
    archivePrefix = "arXiv",
    doi = "10.1103/PhysRevD.72.054022",
    journal = "Phys. Rev. D",
    volume = "72",
    pages = "054022",
    year = "2005"
}

@article{BESIII:2020nbj,
    author = "Ablikim, Medina and others",
    collaboration = "BESIII",
    title = "{Study of Open-Charm Decays and Radiative Transitions of the $X(3872)$}",
    eprint = "2001.01156",
    archivePrefix = "arXiv",
    primaryClass = "hep-ex",
    doi = "10.1103/PhysRevLett.124.242001",
    journal = "Phys. Rev. Lett.",
    volume = "124",
    number = "24",
    pages = "242001",
    year = "2020"
}

@article{Braaten:2019ags,
    author = "Braaten, Eric and He, Li-Ping and Ingles, Kevin",
    title = "{Branching Fractions of the $X(3872)$}",
    eprint = "1908.02807",
    archivePrefix = "arXiv",
    primaryClass = "hep-ph",
    journal = "",
    month = "8",
    year = "2019"
}

@article{Li:2019kpj,
    author = "Li, Chunhua and Yuan, Chang-Zheng",
    title = "{Determination of the absolute branching fractions of $X(3872)$ decays}",
    eprint = "1907.09149",
    archivePrefix = "arXiv",
    primaryClass = "hep-ex",
    doi = "10.1103/PhysRevD.100.094003",
    journal = "Phys. Rev. D",
    volume = "100",
    number = "9",
    pages = "094003",
    year = "2019"
}

@article{Ross:1963lhp,
    author = "Ross, Marc",
    title = "{Position of Resonance Poles Near the Threshold of a Channel}",
    doi = "10.1103/physrevlett.11.450",
    journal = "Phys. Rev. Lett.",
    volume = "11",
    number = "9",
    pages = "450--453",
    year = "1963"
}

@article{Dalitz:1963ek,
    author = "Dalitz, R. H. and Rajasekaran, G.",
    title = "{Resonance poles and mass differences within unitary multiplets}",
    doi = "10.1016/0031-9163(63)90076-3",
    journal = "Phys. Lett.",
    volume = "7",
    pages = "373--377",
    year = "1963"
}

@article{Eden:1963zz,
    author = "Eden, R. J. and Taylor, J. R.",
    title = "{Resonance Multiplets and Broken Symmetry}",
    doi = "10.1103/PhysRevLett.11.516",
    journal = "Phys. Rev. Lett.",
    volume = "11",
    pages = "516--518",
    year = "1963"
}

@article{Pelaez:2026fmu,
    author = "Pel{\'a}ez, Jos{\'e} Ram{\'o}n and Rab{\'a}n, Pablo and de Elvira, Jacobo Ruiz",
    title = "{The unintuitive SU(3) flavor and chiral limits of hadron resonances}",
    eprint = "2606.14634",
    archivePrefix = "arXiv",
    primaryClass = "hep-ph",
    journal = "",
    reportNumber = "IPARCOS-UCM-26-032",
    month = "6",
    year = "2026"
}

@article{Brambilla:2019esw,
    author = "Brambilla, Nora and others",
    title = "{The $XYZ$ states: experimental and theoretical status and perspectives}",
    eprint = "1907.07583",
    archivePrefix = "arXiv",
    primaryClass = "hep-ex",
    doi = "10.1016/j.physrep.2020.05.001",
    journal = "Phys. Rept.",
    volume = "873",
    pages = "1--154",
    year = "2020"
}

@article{Braaten:2004rn,
    author = "Braaten, Eric and Hammer, H.-W.",
    title = "{Universality in few-body systems with large scattering length}",
    eprint = "cond-mat/0410417",
    archivePrefix = "arXiv",
    reportNumber = "INT-PUB-04-27",
    doi = "10.1016/j.physrep.2006.03.001",
    journal = "Phys. Rept.",
    volume = "428",
    pages = "259--390",
    year = "2006"
}

@article{Sadl:2024dbd,
    author = "Sadl, Mitja and Collins, Sara and Guo, Zhi-Hui and Padmanath, M. and Prelovsek, Sasa and Yan, Lin-Wan",
    title = "{Charmoniumlike channels $1^+$ with isospin 1 from lattice and effective field theory}",
    eprint = "2406.09842",
    archivePrefix = "arXiv",
    primaryClass = "hep-lat",
    doi = "10.1103/PhysRevD.111.054513",
    journal = "Phys. Rev. D",
    volume = "111",
    number = "5",
    pages = "054513",
    year = "2025"
}

@article{BESIII:2010dhc,
    author = "Ablikim, M. and others",
    collaboration = "BESIII",
    title = "{Study of $a_0^0(980) - f_0(980)$ mixing}",
    eprint = "1012.5131",
    archivePrefix = "arXiv",
    primaryClass = "hep-ex",
    doi = "10.1103/PhysRevD.83.032003",
    journal = "Phys. Rev. D",
    volume = "83",
    pages = "032003",
    year = "2011"
}

@article{Wu:2011yx,
    author = "Wu, Jia-Jun and Liu, Xiao-Hai and Zhao, Qiang and Zou, Bing-Song",
    title = "{The Puzzle of anomalously large isospin violations in $\eta(1405/1475)\to 3\pi$}",
    eprint = "1108.3772",
    archivePrefix = "arXiv",
    primaryClass = "hep-ph",
    doi = "10.1103/PhysRevLett.108.081803",
    journal = "Phys. Rev. Lett.",
    volume = "108",
    pages = "081803",
    year = "2012"
}

@inproceedings{Manohar:1998xv,
    author = "Manohar, Aneesh V.",
    title = "{Large $N$ QCD}",
    booktitle = "{Les Houches Summer School in Theoretical Physics, Session 68: Probing the Standard Model of Particle Interactions}",
    eprint = "hep-ph/9802419",
    archivePrefix = "arXiv",
    reportNumber = "UCSD-PTH-98-06",
    pages = "1091--1169",
    month = "2",
    year = "1998"
}

@article{Lipkin:1996ny,
    author = "Lipkin, Harry J. and Zou, Bing-Song",
    title = "{Comment on ``When can hadronic loops scuttle the Okubo-Zweig-Iizuka rule?''}",
    doi = "10.1103/PhysRevD.53.6693",
    journal = "Phys. Rev. D",
    volume = "53",
    pages = "6693--6696",
    year = "1996"
}

\end{document}